\documentclass[global,twocolumn]{svjour}

\usepackage{amsmath}
\usepackage{amsfonts}

\usepackage[backend=bibtex,style=numeric-comp,firstinits=true]{biblatex}
\usepackage{mathtools}
\usepackage{etoolbox}
\usepackage{subfig}
\usepackage{floatrow}
\usepackage{rotating}
\usepackage{booktabs}
\usepackage{varwidth}
\usepackage{calc}
\usepackage{tikz}
\usepackage{caption}
\usepackage{xfrac}
\usepackage{nicefrac}
\usepackage{commath}
\usepackage{siunitx}
\usepackage[makeroom]{cancel}

\usetikzlibrary{calc}
\usetikzlibrary{matrix}
\tikzstyle{every picture}+=[remember picture]

\tikzset{
	table/.style={
		matrix of nodes,
		row sep=0,
		column sep=0,
		nodes={rectangle,text width=7ex,align=center},
		text depth=0.25ex,
		text height=4ex,
		font=\footnotesize,
		row 6/.append style={nodes={text height=9ex}},
		column 6/.append style={nodes={text width=4cm}},
		column 7/.append style={nodes={text width=5cm}},
		nodes in empty cells
	}
}

\usepackage{color}	
\definecolor{darkblue}{rgb}{0, 0, 0.5}
\definecolor{darkgreen}{rgb}{0, 0.5, 0}
\definecolor{shaded}{gray}{.6}
\definecolor{matrixbg}{rgb}{1,0.85,0}
\definecolor{matrixfg}{rgb}{0.8,0.6,0}
\definecolor{unknownbg}{rgb}{0.95,0.55,0.48}
\definecolor{unknownfg}{rgb}{0.8,0.25,0.15}
\definecolor{rhsfg}{rgb}{0,0.45,0.75}
\definecolor{rhsbg}{rgb}{0.35,0.70,1}
\definecolor{regfg}{rgb}{0.25,0.55,0.15}
\definecolor{regbg}{rgb}{0.6,0.85,0.5}

\definecolor{offdiagbg}{rgb}{0.95,0.55,0.48}
\definecolor{offdiagfg}{rgb}{0.8,0.25,0.15}
\definecolor{diagbg}{rgb}{1,0.85,0}
\definecolor{diagfg}{rgb}{0.8,0.6,0}

\usepackage{bm}
\let\vec\bm

\let\stdvec\vec
\newcommand{\vecdot}[1]{\dot{\stdvec{#1}}}

\newcommand{\argmin}{\operatornamewithlimits{arg\,min}}
\newcommand{\compl}{~\bot~}

\newcommand{\vectoscalar}[1]{{\renewcommand{\vec}{}#1}}

\newcommand{\vectovecdot}[1]{{\renewcommand{\vec}{\vecdot}#1}}
\newcommand{\vectoscalardot}[1]{{\renewcommand{\vec}{\dot}#1}}

\newcommand{\dt}{\ensuremath{\delta t}}

\newcommand{\setprop}[2]{\ensuremath{\left\{ #1 \,\middle|\, #2 \right\}}}

\newcommand{\intervalCC}[2]{\ensuremath{\left[ #1, #2 \right]}}
\newcommand{\intervalOO}[2]{\ensuremath{\left] #1, #2 \right[}}
\newcommand{\intervalCO}[2]{\ensuremath{\left[ #1, #2 \right[}}

\newcommand{\mat}[1]{{\bf #1}}
\newcommand{\R}{\mathbb{R}}

\newcommand{\dvect}[2]{\ensuremath{\begin{pmatrix}#1\\#2\end{pmatrix}}}

\newcommand{\tvect}[3]{\ensuremath{\begin{pmatrix}#1\\#2\\#3\end{pmatrix}}}
\newcommand{\tvec} [3]{\ensuremath{\begin{pmatrix}#1 & #2 & #3\end{pmatrix}}}

\newcommand{\transp}{{\mathrm{T}}}
\newcommand{\cross}{\times}

\newcommand{\pos}    [1]{\ensuremath{\vec{x}_{#1}}}
\newcommand{\orient} [1]{\ensuremath{\vec{\varphi}_{#1}}}
\newcommand{\linvel} [1]{\ensuremath{\vec{v}_{#1}}}
\newcommand{\angvel} [1]{\ensuremath{\vec{\omega}_{#1}}}
\newcommand{\prelinvel} [1]{\ensuremath{\vec{v}_{#1}^-}}
\newcommand{\preangvel} [1]{\ensuremath{\vec{\omega}_{#1}^-}}

\newcommand{\prepostlinvel} [1]{\ensuremath{\vec{v}_{#1}^{\sfrac{-}{+}}}}
\newcommand{\prepostangvel} [1]{\ensuremath{\vec{\omega}_{#1}^{\sfrac{-}{+}}}}

\newcommand{\mass}   [1]{\ensuremath{m_{#1}}}
\newcommand{\inertia}[1]{\ensuremath{\mat{I}_{#1}}}
\newcommand{\identmat}[1]{\ensuremath{\mat{E}_{#1}}}
\newcommand{\unitvec}[1]{\ensuremath{\vec{e}_{#1}}}

\newcommand{\orientdot}[1]{\ensuremath{\vectovecdot{\orient{#1}}}}

\newcommand{\angveldot}[1]{\ensuremath{\vectovecdot{\angvel{#1}}}}
\newcommand{\prepostlinveldot}[1]{\ensuremath{\vectovecdot{\prepostlinvel{#1}}}}
\newcommand{\prepostangveldot}[1]{\ensuremath{\vectovecdot{\prepostangvel{#1}}}}

\newcommand{\force}    [1]{\ensuremath{\vec{f}_{#1}}}
\newcommand{\torque}   [1]{\ensuremath{\vec{\tau}_{#1}}}
\newcommand{\forceext} [1]{\ensuremath{\vec{f}_{\ifstrequal{#1}{}{ext}{#1,ext}}}}
\newcommand{\torqueext}[1]{\ensuremath{\vec{\tau}_{\ifstrequal{#1}{}{ext}{#1,ext}}}}
\newcommand{\forceextapprox} [1]{\ensuremath{\vec{\tilde{f}}_{\ifstrequal{#1}{}{ext}{#1,ext}}}}
\newcommand{\torqueextapprox}[1]{\ensuremath{\vec{\tilde{\tau}}_{\ifstrequal{#1}{}{ext}{#1,ext}}}}
\newcommand{\linimp}   [1]{\ensuremath{\vec{\Delta p}_{#1}}}
\newcommand{\angimp}   [1]{\ensuremath{\vec{\Delta L}_{#1}}}

\newcommand{\angmom}   [1]{\ensuremath{\vec{L}_{#1}}}
\newcommand{\tInt}{\ensuremath{t_p}}
\newcommand{\tImp}{\ensuremath{t_q}}

\newcommand{\cof}[1]{\ensuremath{\mu_{#1}}}

\newcommand{\contactforce}[1]{\ensuremath{\vec{\lambda}_{#1}}}
\newcommand{\contactforceCFn}[1]{\ensuremath{\vectoscalar{\contactforce{\ifstrequal{#1}{}{n}{#1,n}}}}}
\newcommand{\contactforceCFto}[1]{\ensuremath{\contactforce{\ifstrequal{#1}{}{to}{#1,to}}}}

\newcommand{\contactforceapprox}[1]{\ensuremath{\vec{\tilde{\lambda}}_{#1}}}
\newcommand{\contactforceapproxCFn}[1]{\ensuremath{\vectoscalar{\contactforceapprox{\ifstrequal{#1}{}{n}{#1,n}}}}}
\newcommand{\contactforceapproxCFto}[1]{\ensuremath{\contactforceapprox{\ifstrequal{#1}{}{to}{#1,to}}}}

\newcommand{\contactimpulse}[1]{\ensuremath{\vec{\Lambda}_{#1}}}

\newcommand{\contactimpulseCFn}[1]{\ensuremath{\vectoscalar{\contactimpulse{\ifstrequal{#1}{}{n}{#1,n}}}}}
\newcommand{\contactimpulseCFto}[1]{\ensuremath{\contactimpulse{\ifstrequal{#1}{}{to}{#1,to}}}}

\newcommand{\relvelWF}[1]{\ensuremath{\vec{\delta v}_{#1}}}
\newcommand{\relvelCF}[1]{\ensuremath{\vec{\delta v}_{#1}}}
\newcommand{\relvelCFn}[1]{\ensuremath{\vectoscalar{\relvelCF{\ifstrequal{#1}{}{n}{#1,n}}}}}
\newcommand{\relvelCFt}[1]{\ensuremath{\vectoscalar{\relvelCF{\ifstrequal{#1}{}{t}{#1,t}}}}}
\newcommand{\relvelCFo}[1]{\ensuremath{\vectoscalar{\relvelCF{\ifstrequal{#1}{}{o}{#1,o}}}}}
\newcommand{\relvelCFto}[1]{\ensuremath{\relvelCF{\ifstrequal{#1}{}{to}{#1,to}}}}

\newcommand{\discreterelvel}[1]{\ensuremath{{\vec{\delta v}_{#1}'}}}
\newcommand{\discreterelvelCFn}[1]{\ensuremath{\vectoscalar{\discreterelvel{\ifstrequal{#1}{}{n}{#1,n}}}}}
\newcommand{\discreterelvelCFto}[1]{\ensuremath{\discreterelvel{\ifstrequal{#1}{}{to}{#1,to}}}}

\newcommand{\prerelvel}[1]{\ensuremath{{\vec{\delta v}_{#1}^-}}}
\newcommand{\prerelvelCFn}[1]{\ensuremath{\vectoscalar{\prerelvel{\ifstrequal{#1}{}{n}{#1,n}}}}}
\newcommand{\prerelveldotCFn}[1]{\ensuremath{\vectoscalardot{\prerelvel{\ifstrequal{#1}{}{n}{#1,n}}}}}

\newcommand{\prerelvelCFto}[1]{\ensuremath{\prerelvel{\ifstrequal{#1}{}{to}{#1,to}}}}

\newcommand{\postrelvel}[1]{\ensuremath{{\vec{\delta v}_{#1}^+}}}
\newcommand{\postrelvelCFn}[1]{\ensuremath{\vectoscalar{\postrelvel{\ifstrequal{#1}{}{n}{#1,n}}}}}

\newcommand{\postrelvelCFto}[1]{\ensuremath{\postrelvel{\ifstrequal{#1}{}{to}{#1,to}}}}
\newcommand{\postrelveldotCFto}[1]{\ensuremath{\vectovecdot{\postrelvelCFto{#1}}}}
\newcommand{\prepostrelvel}[1]{\ensuremath{{\vec{\delta v}_{#1}^{\sfrac{-}{+}}}}}
\newcommand{\prepostrelveldot}[1]{\ensuremath{\vectovecdot{\prepostrelvel{#1}}}}
\newcommand{\prepostrelveldotCFn}[1]{\ensuremath{\vectoscalardot{\prepostrelvel{\ifstrequal{#1}{}{n}{#1,n}}}}}
\newcommand{\prepostrelvelCFn}[1]{\ensuremath{\vectoscalar{\prepostrelvel{\ifstrequal{#1}{}{n}{#1,n}}}}}
\newcommand{\prepostrelvelCFto}[1]{\ensuremath{\prepostrelvel{\ifstrequal{#1}{}{to}{#1,to}}}}
\newcommand{\primerelvel}[1]{\ensuremath{{\vec{\delta v}_{#1}'}}}
\newcommand{\primerelvelCFn}[1]{\ensuremath{\vectoscalar{\primerelvel{\ifstrequal{#1}{}{n}{#1,n}}}}}

\newcommand{\primerelvelCFto}[1]{\ensuremath{\primerelvel{\ifstrequal{#1}{}{to}{#1,to}}}}

\newcommand{\contactpos}    [1]{\ensuremath{\vec{\hat{x}}_{#1}}}

\newcommand{\diag}{\operatornamewithlimits{diag}}
\newcommand{\vertcat}{\operatornamewithlimits{vertcat}}
\newcommand{\numbodies}{\ensuremath{\nu_b}}
\newcommand{\numcontacts}{\ensuremath{\nu_c}}

\newcommand{\shape}    [1]{\ensuremath{\mathcal{S}_{#1}}}

\newcommand*\diff{\mathop{}\!\mathrm{d}}
\newcommand{\atan}[1]{\operatorname{atan#1}}
\newcommand{\acos}{\operatorname{acos}}

\newcommand{\twodots}{\ensuremath{\mathrel{\ldotp\ldotp}}}

\DeclareMathAlphabet{\mathpzc}{OT1}{pzc}{m}{it}

\renewcommand{\figref}[1]{Fig.~\ref{#1}}
\newcommand{\shrinkeqnnew}[2]{\resizebox{#1\linewidth}{!}{\begin{varwidth}[t]{2\linewidth}\ensuremath{\displaystyle{#2}}\end{varwidth}}}

\newcounter{inlinesubeq}[equation]
\makeatletter
\newcommand{\inlinesubeqno}{\refstepcounter{inlinesubeq}\def\@currentlabel{\theequation \alph{inlinesubeq}}(\@currentlabel)}
\makeatother

\title{The Maximum Dissipation Principle in Rigid-Body Dynamics with Purely Inelastic Impacts}
\author{Tobias Preclik\inst{1} \and Sebastian Eibl\inst{1} \and Ulrich R\"{u}de\inst{1,2}}
\mail{Tobias Preclik\\tobias.preclik@fau.de}
\institute{Lehrstuhl f\"{u}r Informatik 10 (Systemsimulation), Friedrich-Alexander Universit\"{a}t Erlangen-N\"{u}rnberg, Cauerstr.~11, 91058 Erlangen, Germany \and
CERFACS, 42 Avenue Gaspard Coriolis, 31057 Toulouse, Cedex 01, France}

\begin{document}
	\maketitle

	\begin{abstract}
	Formulating a consistent theory for rigid-body dynamics with impacts is an intricate problem. Twenty years ago
	Stewart published the first consistent theory with purely inelastic impacts and an impulsive friction model
	analogous to Coulomb friction. In this paper we demonstrate that the consistent impact model can exhibit multiple solutions
	with a varying degree of dissipation even in the single-contact case.
	Replacing the impulsive friction model based on Coulomb friction by a model based on the maximum dissipation
	principle resolves the non-uniqueness in the single-contact impact problem. The paper constructs the alternative
	impact model and presents integral equations describing rigid-body dynamics with a non-impulsive and non-compliant
	contact model and an associated purely inelastic impact model maximizing dissipation. An analytic solution is derived
	for the single-contact impact problem. The models are then embedded into a time-stepping scheme.
	The macroscopic behaviour is compared to Coulomb friction in a large-scale granular flow problem.
	\end{abstract}


	\newpage

	\section{Introduction}
		
		Simulating mechanical systems on computers requires a model for describing the dynamics of the mechanical parts.
		Models that can describe the deformation of the mechanical parts require a high number of parameters to
		describe the deformation. If the core of the mechanical parts can be assumed to not deform under the considered
		loads, the parameters describing the state of a mechanical part can be reduced to that of a rigid body: An
		invariant shape with an associated mass and principal moments of inertia as well as the spatial orientation, position, linear and angular velocity of
		the shape. The interaction of multiple such mechanical parts must be described by another model determining the
		dynamics of the mechanical parts in contact, where the contact model usually allows compliance in a localized
		contact region. If mechanical parts collide, the contact dynamics typically occur on a time-scale that is
		significantly smaller than that of the motions between successive collisions. Resolving each such collision
		micro-dynamics in computer simulations can become computationally expensive.
		Alternatively, the relation between the pre- and post-collision state variables can be described by an impact
		model~\cite{stronge04}. By condensing the impact dynamics to a single point in time, only the response
		of the relative contact velocities has to be specified. In order to instantaneously turn a colliding state
		into a non-colliding state a contact reaction impulse must be applied, so that the necessary
		discontinuities in the velocities can be effected. The impact model must then be combined with a non-compliant
		contact model in order to determine the contact reaction forces and contact reaction impulses.
		Non-compliant contact models alone cannot resolve collisions. For non-compliant contact models with Coulomb
		friction even non-colliding contact situations exist~(shocks) where an impulse becomes necessary to resolve
		the contact~\cite{stewart00}. These paradoxical situations were first published by Painlev\'{e}~\cite{painleve1895}.
		The system including the non-compliant contact and impact models are often mathematically described in terms of
		measure differential inclusions~(MDI)~\cite{moreau88}.

		The rigid body simplification combined with the impact simplification considerably reduce the computational
		burden in simulations with many mechanical parts. Applications range from robotics~\cite{nuseirat00,jia13},
		virtual reality~\cite{sauer98}, physics-based animation~\cite{erleben04} to granular matter simulations~\cite{tasora10}.
		In particular, granular dynamics can require a very large number of particles and are insufficiently understood
		to date with and without an interstitial fluid phase~\cite{mitarai12}. Optimizing mechanical devices like
		powder mixers~\cite{hassanpour11} or grinding mills~\cite{mishra92,jayasundara11} is of economical importance.
		Powder mixing is important in detergent, cosmetic, food and pharmaceutical manufacturing, to name just a few applications. Understanding
		granular matter is also of crucial importance for safety reasons: Assessing the stability of slopes is
		important to prevent rock slides, land slides and snow avalanches and getting the particle distribution right
		in pebble-bed nuclear reactors is important to guarantee safe and performant operation~\cite{tasora10}.

		The construction of an impact model that in combination with a non-compliant contact model leads to
		a consistent theory for rigid body dynamics is non-trivial. Ideally, the solution of an impact model
		for a collision is a limit point of a sequence of solutions of the collision based on a compliant contact
		model with increasing stiffness. The sequence of solutions is uniquely determined and the increasing
		stiffness decreases the duration of the collision towards an instantaneous event. The solution of
		the collision based on a compliant contact model corresponds to the integral of the contact reaction
		forces over the collision duration. Using such an approach Stronge constructs an energetically consistent restitution
		hypothesis in \cite{stronge90}, Mirtich solves rigid-body dynamics with impacts for virtual reality applications,
		where permanent contacts are treated as sequences of collisions~\cite{mirtich95, mirtich96}.
		And lately, Jia and Wang showed in \cite{jia16} how contact reaction impulses can be computed from the limit of a
		contact model with linear normal stiffness and Coulomb friction for general collisions
		(central or eccentric, direct or oblique) in three dimensions. The authors established a condition
		that, if met, guarantees solution existence. Whether solutions exist unconditionally
		and whether the impact model in combination with a non-compliant contact model leads to
		a consistent theory remain open problems.

		However, other impact models exist that possess solutions unconditionally. Stewart showed
		in~\cite{stewart98} for a time-stepping scheme based on \cite{anitescu97},
		that it converges to a solution of an MDI as the time-step
		size decreases. Recently, Gavrea et al. extended the result to systems including joints
		in~\cite{gavrea08}. The
		MDI describes rigid-body dynamics with Coulomb friction and purely inelastic impacts (collisions and shocks),
		where the frictional impulses are required to directly oppose the \emph{post-impulse}
		relative contact velocities in the tangential planes thus imitating Coulomb's
		friction law in the case of impulses. A consequence of this is that Stewart proved
		that solutions exist for the MDI and thus resolved paradoxical
		configurations in rigid-body dynamics with non-compliant contacts and Coulomb friction,
		where apparently no solutions exist even though no collisions are present. Stewart made no attempt to show uniqueness
		of solutions. In fact in section \ref{sec:non-unique} we present an example demonstrating the
		existence of multiple solutions of a numerically constructed single-contact impact problem.
		The non-uniqueness is in this single-contact case directly related to the choice of the frictional impact model.
		In this paper we construct an alternative frictional impact model having a
		unique solution in the single-contact case. The friction model is based on the
		maximum dissipation principle~\cite{stewart00} and takes into account the coupling between the normal component
		and the tangential components of the contact reactions. The non-uniqueness
		due to redundant constraints in the multi-contact case remains unaffected~\cite{popa15}
		as well as non-uniqueness in the non-compliant contact model.
		
		In section~\ref{sec:continuous} we present integral equations describing
		rigid-body dynamics with impact and friction, where the Coulomb friction model
		on the impulsive reactions is replaced by a friction model based on the
		maximum dissipation principle. In section~\ref{sec:numerics} the model
		is embedded into an impulse-velocity time-stepping scheme for numerically
		integrating multi-contact problems and an analytic solution of the
		single-contact problem is established. Subsequently, section~\ref{sec:results}
		presents results for single-contact problems and the
		macro-scale behaviour of the friction model in the simulation of a
		large-scale granular flow problem. The paper summarizes the results and concludes in
		section~\ref{sec:summary}.

	\section{Continuous System}
	\label{sec:continuous}

		Each particle~$i$ is associated with a co-rotating body frame. The
		origin of the body frame corresponds to the center of mass of the
		particle. Let $\pos{i}$ be the position function of the body frame in
		the inertial frame. The position function is non-smooth when impulses
		act. The derivative with respect to time is the discontinuous linear
		velocity function $\linvel{i}$ with left- and
		right-limits $\linvel{i}^-$ and $\linvel{i}^+$. The orientation of the body frame in the
		inertial frame can be represented by a unit quaternion. Instead of
		mixing vector and quaternion algebra a quaternion
		$q_w + q_x \mathrm{i} + q_y \mathrm{j} + q_z \mathrm{k} \in \mathbb{H}$
		describing the orientation of the body frame of particle~$i$ at time~$t$
		is represented as a vector $\orient{i}(t) = (q_w, q_x, q_y, q_z)^\transp \in \R^4$. The orientation function \orient{i} is non-smooth and the
		left- and right limit of the derivative at time~$t$ is then~\cite{diebel06}
		\begin{equation*}
			\shrinkeqnnew{1}{
			\orientdot{i}^{\sfrac{-}{+}}(t) = \frac{1}{2} \begin{bmatrix} -q_x & -q_y & -q_z \\ q_w & q_z & -q_y \\ -q_z & q_w & q_x \\ q_y & -q_x & q_w \end{bmatrix} \angvel{i}^{\sfrac{-}{+}}(t) = \mat{Q}_{ii}(\orient{i}(t)) \angvel{i}^{\sfrac{-}{+}}(t),
			}
		\end{equation*}
		where~$\angvel{i}^{\sfrac{-}{+}}(t) \in \R^3$ is the angular velocity of the
		particle \sfrac{before}{after} applying impulses at time~$t$.
		We also introduce the quaternion matrix function~$\mat{Q}_{ii}$ for
		abbreviating the notation.
		
		The mass of the particle is denoted by~\mass{i} and is invariant with
		respect to time. The inertia tensor of the particle in the inertial
		frame changes with respect to the orientation of the body frame. It can
		be expressed in terms of the constant inertia tensor in the body
		frame~$\inertia{i,0}$.
		For time~$t$ the inertia tensor in the inertial frame is given by
		\begin{equation*}
			\inertia{ii}(\orient{i}(t)) = \mat{R}(\orient{i}(t)) \inertia{i,0} \mat{R}(\orient{i}(t))^\transp,
		\end{equation*}
		where $\mat{R}(\orient{i}(t))$ is the rotation matrix corresponding to
		the orientation $\orient{i}(t)$. The rotation matrix changes the basis
		from the particle's body frame to the inertial frame. Choosing the body
		frame such that the axes match the principal axes of the particle, the
		body frame inertia tensor~$\inertia{i,0}$ can be enforced to be
		diagonal.

		In a system with \numbodies{} particles, let
		$\pos{}(t)$~($\orient{}(t)$, $\linvel{}^{\sfrac{-}{+}}(t)$, and $\angvel{}^{\sfrac{-}{+}}(t)$) be the
		vertical concatenation of all particles' positions (orientations, linear velocities, and angular velocities) at time~$t$:
		\begin{equation*}
			\pos{}(t) = \vertcat_{i=1\twodots\numbodies} \pos{i}(t) \in \R^{3\numbodies{}}.
		\end{equation*}
		Let
		\begin{equation*}
			\begin{split}
				\mat{Q}(\orient{}(t)) & = \diag_{i=1\twodots\numbodies} \mat{Q}_{ii}(\orient{i}(t)), \\
				\inertia{}(\orient{}(t)) & = \diag_{i=1\twodots\numbodies} \inertia{ii}(\orient{i}(t)),
			\end{split}
		\end{equation*}
		and let $\mat{M}(\orient{}(t))$ be the
		block-diagonal mass matrix containing
		$\diag_{i=1\twodots\numbodies} \mass{i} \identmat{3}$ in the upper-left
		quadrant and $\inertia{}(\orient{}(t))$ in the lower-right quadrant,
		where $\identmat{3}$ is the $3 \times 3$~identity matrix. Then given
		initial conditions at time~$t_0$, the state of the system at time~$t$
		is described by the integral equations
		\begin{alignat*}{2}
			&& \dvect{\pos{}(t)}{\orient{}(t)} & = \dvect{\pos{}(t_0)}{\orient{}(t_0)} + \int_{t_0}^t \dvect{\linvel{}^-(\tInt)}{\mat{Q}(\orient{}(\tInt)) \angvel{}^-(\tInt)} \diff \tInt, \\
			&& \dvect{\linvel{}^{\sfrac{-}{+}}(t)}{\angvel{}^{\sfrac{-}{+}}(t)} & = \dvect{\linvel{}^-(t_0)}{\angvel{}^-(t_0)}
			   + \sum_{\crampedclap{\substack{\tImp \in \mathcal{T}_q\\t_0\,\leq\,\tImp\,\nicefrac{<}{\leq}\,t}}} \mat{M}(\orient{}(\tImp))^{-1} \dvect{ \linimp{}(\tImp)}{\angimp{}(\tImp)} \\
			&\rlap{\resizebox{\linewidth}{!}{$\displaystyle + \int_{t_0}^t \mat{M}(\orient{}(\tInt))^{-1} \dvect{ \force{}(\tInt)}{\torque{}(\tInt) - \angvel{}^-(\tInt) \cross \inertia{}(\orient{}(\tInt)) \angvel{}^-(\tInt)} \diff \tInt,$}} &&
		\end{alignat*}
		%
		%
		where $\mathcal{T}_q$ is the set containing all points in time \tImp{},
		where impulses are present, that is (linear)
		impulse~$\linimp{}(\tImp{}) \in \R^{3\numbodies{}}$ or angular
		impulse~$\angimp{}(\tImp{}) \in \R^{3\numbodies{}}$ is non-zero:
		\begin{equation*}
			\mathcal{T}_q = \setprop{t}{\linimp{}(t) \neq \vec 0 \lor \angimp{}(t) \neq \vec 0}.
		\end{equation*}
		The terms $\force{}(t) \in \R^{3\numbodies}$ and
		$\torque{}(t) \in \R^{3\numbodies}$ are the forces and torques acting
		on the particles. Note that the inverse of the mass matrix always
		exists, since it is symmetric positive-definite~(SPD) - a property which it inherits
		from its diagonal blocks. The cross-product term
		is to be understood as the vertical concatenation of all single-particle
		cross-products:
		\begin{equation*}
			\angvel{}^-(t) \cross \inertia{}(\orient{}(t)) \angvel{}^-(t) = \vertcat_{i=1\twodots\numbodies} \angvel{i}^-(t) \cross \inertia{ii}(\orient{i}(t)) \angvel{i}^-(t).
		\end{equation*}
		The appearance of the term stems from the fact that the torque function
		$\torque{i}$ corresponds to the time-derivative of the angular
		momentum function~$\angmom{i}$ (for non-impulsive points in time), which in turn
		is the product of the time-varying inertia tensor and the angular
		velocity. Hence,
		\begin{align*}
			\torque{i}(\tInt) & = \od{\angmom{i}(t)}{t}\sVert[2]_{t=\tInt} = \od{}{t} \inertia{ii}(\orient{i}(t)) \angvel{i}(t) \sVert[2]_{t=\tInt} \\
				              & = \inertia{ii}(\orient{i}(\tInt)) \angveldot{i}(\tInt) + \od{}{t} \inertia{ii}(\orient{i}(t)) \sVert[2]_{t=\tInt} \angvel{i}(\tInt) \\
			                  & = \inertia{ii}(\orient{i}(\tInt)) \angveldot{i}(\tInt) + \angvel{i}(\tInt) \cross \inertia{ii}(\orient{i}(\tInt)) \angvel{i}(\tInt).
		\end{align*}


		The forces, torques, linear impulses, and angular impulses at time~$t$
		include components from non-impulsive contact reactions
		$\contactforce{}(t) \in \R^{3\numcontacts}$ and impulsive contact
		reactions $\contactimpulse{}(t) \in \R^{3\numcontacts}$, where
		\numcontacts{} is the number of contacts in the particle system.
		Each contact~$j$ involves a pair of particles $(i_1(j)$, $i_2(j))$.
		By convention let contact reactions act positively on the first
		particle~$i_1(j)$ and negatively on the second particle~$i_2(j)$.
		Each contact~$j$ is also associated with a contact frame. Let the first
		axis of the contact frame correspond to the contact normal $\vec n_j(t)$ pointing from
		particle~$i_2(j)$ towards particle~$i_1(j)$ by convention, and let
		orthonormal vectors $\vec t_j(t)$, and $\vec o_j(t)$ complete the contact
		frame. Let $\contactpos{j}(t)$ denote the position of the contact frame
		in the inertial frame. Then, subsuming all forces and torques on
		particle~$i$, which are not due to contact reactions, as external
		forces~$\forceext{i}(t)$ and external torques~$\torqueext{i}(t)$, the equations
		\begin{align*}
			\force{i}(t) & = \forceext{i}(t) && + \sum_{\substack{j = 1\twodots\numcontacts\\i_1(j)=i}} \contactforce{j}(t) - \sum_{\substack{j = 1\twodots\numcontacts\\i_2(j)=i}} \contactforce{j}(t), \\
			\torque{i}(t) & = \torqueext{i}(t) && + \sum_{\substack{j = 1\twodots\numcontacts\\i_1(j)=i}} (\contactpos{j}(t) - \pos{i}(t)) \cross \contactforce{j}(t) \\
			&&& - \sum_{\substack{j = 1\twodots\numcontacts\\i_2(j)=i}} (\contactpos{j}(t) - \pos{i}(t)) \cross \contactforce{j}(t),
		\end{align*}
		define a wrench matrix function~$\mat{W}$ relating the wrenches to the contact reactions. This relation extends to impulsive reactions and linear and angular impulses:
		\begin{equation*}
			\begin{split}
				\dvect{\force{}(t)}{\torque{}(t)} & = \dvect{\forceext{}(t)}{\torqueext{}(t)} + \mat{W}(t) \contactforce{}(t),\\
				\dvect{\linimp{}(t)}{\angimp{}(t)} & = \dvect{\linimp{ext}(t)}{\angimp{ext}(t)} + \mat{W}(t) \contactimpulse{}(t).
			\end{split}
		\end{equation*}
		The impulsive and non-impulsive contact reactions are then given
		implicitly as solutions of contact constraints. The contact constraints
		are usually non-linear and underdetermined depending on the specific
		contact model employed.

		The formulation of the contact constraints requires the
		rigorous introduction of the contact position function~$\contactpos{}$,
		the contact normal function~$\vec n$, the signed contact distance
		function~$\xi$ and the relative contact velocity function~$\relvelWF{}$.
		The latter is straightforward and for a contact~$j$ given by
		\begin{equation*}
			\begin{split}
				\relvelWF{j}^{\sfrac{-}{+}}(t) & = \linvel{i_1(j)}^{\sfrac{-}{+}}(t) + \angvel{i_1(j)}^{\sfrac{-}{+}}(t) \cross (\contactpos{j}(t) - \pos{i_1(j)}(t)) \\
				                               & - \linvel{i_2(j)}^{\sfrac{-}{+}}(t) - \angvel{i_2(j)}^{\sfrac{-}{+}}(t) \cross (\contactpos{j}(t) - \pos{i_2(j)}(t)).
			\end{split}
		\end{equation*}
		It can be shown that
		\begin{equation*}
			\relvelWF{}^{\sfrac{-}{+}}(t) = \mat{W}(t)^\transp \dvect{\linvel{}^{\sfrac{-}{+}}(t)}{\angvel{}^{\sfrac{-}{+}}(t)}.
		\end{equation*}

		The definition of the other three functions are difficult to state in
		sufficient generality. We confine ourselves here to
		definitions that are at least well-defined for spherical particles.
		Let $\shape{i}(t)$ be the set of points in the inertial frame
		defining the shape of particle~$i$ at time~$t$, and let
		$f_{\shape{}}: \R^3 \rightarrow \R$ be the signed distance function associated
		with the shape~\shape{}. The signed distance function shall be negative in the
		interior of the shape. Then, let
		\begin{equation*}
			\contactpos{j}(t) \in \argmin_{f_{\shape{i_2(j)}(t)}(\vec y) \leq 0} f_{\shape{i_1(j)}(t)}(\vec y)
		\end{equation*}
		be the contact point between the pair of particles $(i_1(j),\allowbreak i_2(j))$.
		If the boundary of the shape is sufficiently smooth and the overlap
		sufficiently small, the contact position is uniquely determined and the
		gradient of the signed distance function exists. Then the contact normal
		is given by
		\begin{equation*}
			\vec n_j(t) = \nabla f_{\shape{i_2(j)}(t)}(\contactpos{j}(t)).
		\end{equation*}
		The signed contact distance function is then simply
		\begin{equation*}
			\xi_j(t) = f_{\shape{i_1(j)}(t)}(\contactpos{j}(t)).
		\end{equation*}
		These specific definitions of the contact functions limit the number of contacts~\numcontacts{}
		to the number of particle pairs~$\frac{\numbodies}{2}(\numbodies - 1)$.
		To simplify the description of the contact constraints,
		subscript~$n$ denotes the projection of a vector to the contact
		normal~(e.g.\ $\contactforceCFn{j}(t) := \vec n_j(t)^\transp \contactforce{j}(t) \in \R$)
		and subscript~$to$ denotes the vector of projections of a vector to
		the contact tangential and contact
		orthogonal~(e.g.\ $\contactforceCFto{j}(t) := (\vec t_j(t)^\transp \contactforce{j}(t),\ \vec o_j(t)^\transp \contactforce{j}(t))^\transp \in \R^2$).

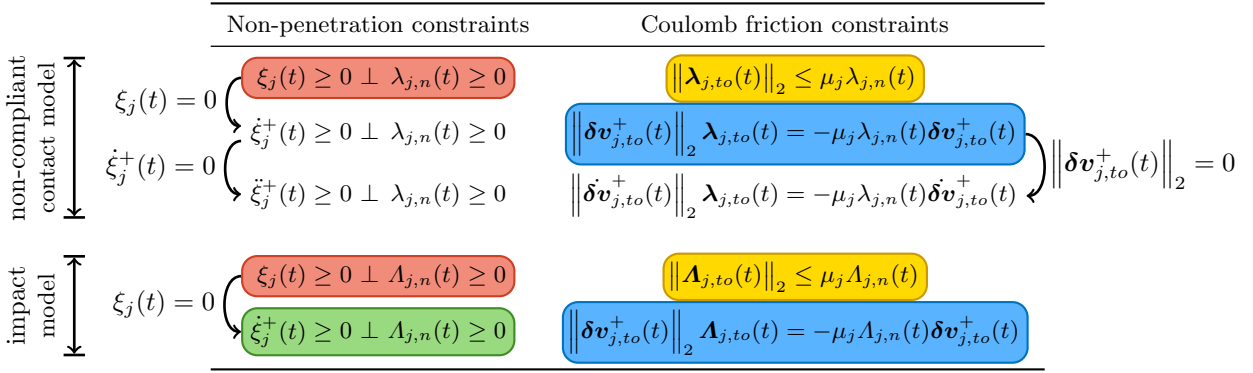
\begin{figure*}
			\begin{center}
				\begin{tabular}{ccc}
					\toprule
					Non-penetration constraints & Coulomb friction constraints \\
					\midrule
					\tikz[baseline]{\node[draw=offdiagfg,fill=offdiagbg,thick,rounded corners=2mm,anchor=base] (t1)  {$\displaystyle \parbox{\widthof{$\displaystyle \ddot{\xi}_j^+(t)$}}{\hfill $\displaystyle \xi_j(t)$} \geq 0 \compl \parbox{\widthof{$\displaystyle \contactimpulseCFn{j}(t)$}}{\hfill $\displaystyle \contactforceCFn{j}(t)$} \geq 0$};} &	
					\tikz[baseline]{\node[draw=diagfg,fill=diagbg,      thick,rounded corners=2mm,anchor=base] (t2)  {$\displaystyle \norm{\contactforceCFto{j}(t)}_2 \leq  \cof{j} \contactforceCFn{j}(t)$};} \\	
					\tikz[baseline]{\node[                              thick,rounded corners=2mm,anchor=base] (t3)  {$\displaystyle \parbox{\widthof{$\displaystyle \ddot{\xi}_j^+(t)$}}{\hfill $\displaystyle \dot{\xi}_j^+(t)$} \geq 0 \compl \parbox{\widthof{$\displaystyle \contactimpulseCFn{j}(t)$}}{\hfill $\displaystyle \contactforceCFn{j}(t)$} \geq 0$};} &
					\tikz[baseline]{\node[draw=rhsfg,fill=rhsbg,        thick,rounded corners=2mm,anchor=base] (t4)  {$\displaystyle \norm{\postrelvelCFto{j}(t)}_2 \contactforceCFto{j}(t) = -\cof{j} \contactforceCFn{j}(t) \postrelvelCFto{j}(t)$};} \\
					\tikz[baseline]{\node[                              thick,rounded corners=2mm,anchor=base] (t5)  {$\displaystyle \ddot{\xi}_j^+(t) \geq 0 \compl \parbox{\widthof{$\displaystyle \contactimpulseCFn{j}(t)$}}{\hfill $\displaystyle \contactforceCFn{j}(t)$} \geq 0$};} &	
					\tikz[baseline]{\node[                              thick,rounded corners=2mm,anchor=base] (t6)  {$\displaystyle \norm{\postrelveldotCFto{j}(t)}_2 \contactforceCFto{j}(t) = -\cof{j} \contactforceCFn{j}(t) \postrelveldotCFto{j}(t)$};} \\
					\addlinespace
					\addlinespace
					\tikz[baseline]{\node[draw=offdiagfg,fill=offdiagbg,thick,rounded corners=2mm,anchor=base] (t7)  {$\displaystyle \parbox{\widthof{$\displaystyle \ddot{\xi}_j^+(t)$}}{\hfill $\displaystyle \xi_j(t)$} \geq 0 \compl \parbox{\widthof{$\contactimpulseCFn{j}(t)$}}{\hfill$\contactimpulseCFn{j}(t)$} \geq 0$};} &	
					\tikz[baseline]{\node[draw=diagfg,fill=diagbg,      thick,rounded corners=2mm,anchor=base] (t8)  {$\displaystyle \norm{\contactimpulseCFto{j}(t)}_2 \leq \cof{j} \contactimpulseCFn{j}(t)$};} \\	
					\tikz[baseline]{\node[draw=regfg,fill=regbg,        thick,rounded corners=2mm,anchor=base] (t9)  {$\displaystyle \parbox{\widthof{$\displaystyle \ddot{\xi}_j^+(t)$}}{\hfill $\displaystyle \dot{\xi}_j^+(t)$} \geq 0 \compl \parbox{\widthof{$\contactimpulseCFn{j}(t)$}}{\hfill$\contactimpulseCFn{j}(t)$} \geq 0$};} &	
					\tikz[baseline]{\node[draw=rhsfg,fill=rhsbg,        thick,rounded corners=2mm,anchor=base] (t10) {$\displaystyle \norm{\postrelvelCFto{j}(t)}_2 \contactimpulseCFto{j}(t) = -\cof{j} \contactimpulseCFn{j}(t) \postrelvelCFto{j}(t)$};} \\
					\bottomrule
				\end{tabular}
			\end{center}
			
			\begin{tikzpicture}[overlay]
				\path[->,   line width=1pt] (t1)                         edge [out=180, in=180] node [auto,swap] {\normalsize $\xi_j(t) = 0$}                        ($(t3.west)+(0mm,1mm)$);
				\path[->,   line width=1pt] ($(t3.west)+(0mm,-1mm)$)     edge [out=180, in=180] node [auto,swap] {\normalsize $\dot{\xi}_j^+(t) = 0$}                (t5);
				\path[->,   line width=1pt] (t7)                         edge [out=180, in=180] node [auto,swap] {\normalsize $\xi_j(t) = 0$}                        (t9);
				\path[->,   line width=1pt] (t4)                         edge [out=0,   in=0]   node [auto]      {\normalsize $\norm{\postrelvelCFto{j}(t)}_2 = 0$}  (t6);
				\path[|<->|,line width=1pt] ($(t5.south)+(-4.0cm,0cm)$)  edge                   node [sloped,align=center,above=1ex] {\normalsize non-compliant\\ contact model} ($(t1.north)+(-4.0cm,0cm)$);
				\path[|<->|,line width=1pt] ($(t9.south)+(-4.0cm,0cm)$)  edge                   node [sloped,align=center,above=1ex] {\normalsize impact\\ model}                ($(t7.north)+(-4.0cm,0cm)$);
			\end{tikzpicture}

			\caption[]{
				\tikz \draw[offdiagfg,fill=offdiagbg,thick] (0,0) circle (0.7ex);\,Signorini condition,
				\tikz \draw[regfg,fill=regbg,thick] (0,0) circle (0.7ex);\,restitution hypothesis,
				\tikz \draw[diagfg,fill=diagbg,thick] (0,0) circle (0.7ex);\,friction cone condition,
				\tikz \draw[rhsfg,fill=rhsbg,thick] (0,0) circle (0.7ex);\,frictional reaction opposes slip
			}

			\label{fig:contact_constraints}
		\end{figure*}

		Then the contact constraints for an inelastic contact with Coulomb friction are
		listed in \figref{fig:contact_constraints}.
		The constraints can be classified
		into non-penetration constraints and Coulomb friction constraints. Both classes
		can be subdivided into impulsive and non-impulsive constraints. Impulsive constraints
		determine impulsive contact reactions~$\contactimpulse{}(t)$ and non-impulsive
		constraints determine non-impulsive contact reactions~$\contactforce{}(t)$.	Some constraints
		are understood to be enabled only if a precondition holds. For instance
		the restitution hypothesis should only constrain the solution if the contact is
		closed~($\xi_j(t) = 0$). This precondition is indicated by an arrow.
		The arrow originates from a constraint, which enables the constraint
		if the precondition becomes active. In the case of the restitution hypothesis the
		contact needs to close first and the arrow thus originates from the
		impulsive Signorini condition. The Signorini condition ensures that
		contact reactions are non-negative (non-adhesive) if the contact is
		closed and are zero if the contact is open. This relation is
		expressed by the complementarity condition~$\compl$ and the corresponding inequalities.
		A similar chain of non-penetration constraints exists for the
		non-impulsive contact reactions. However, in that chain the contact
		reaction in the worst case can only be determined after the constraint
		on the acceleration level became enabled.
		
		The impulsive and non-impulsive Coulomb friction constraints require
		the contact reaction to reside within a friction cone. The coefficient of
		friction~$\cof{j}$ determines the aperture $2 \tan^{-1} \cof{j}$
		of the cone. The cone is aligned along the contact normal. The friction
		cone condition limits the Euclidean norm of the frictional reaction by an
		upper bound proportional to the contact reaction in normal direction.
		The direction of the frictional reaction is required to oppose the
		relative tangential contact velocity in the case of a sliding~(dynamic)
		contact and its Euclidean norm must be at its limit. This is expressed in
		the velocity-level equation. In the case of a sticking~(static)
		contact, the velocity-level equation is universally valid. However, the
		zero slip enables the acceleration-level constraint. Then, the
		direction of the frictional reaction is required to oppose the relative
		tangential contact acceleration.

		The work performed by the frictional contact reaction force of
		contact~$j$ for a non-impulsive time
		span~\intervalCC{\underline{t}}{\overline{t}} is
		\begin{equation*}
			\int_{\underline{t}}^{\overline{t}} \contactforceCFto{j}(t)^\transp \postrelvelCFto{j}(t) \diff t,
		\end{equation*}
		where the Coulomb friction force performs no work to the extent possible
		as expressed in the acceleration-level constraint. However, if sliding is
		inevitable, the Coulomb friction force maximizes dissipation by
		minimizing the integrand through the velocity-level constraint, which
		requires the friction force to directly oppose the relative contact
		velocity. The velocity-level Coulomb constraint can be formulated
		equivalently using the \emph{maximum dissipation principle} as pointed
		out by Stewart in his review paper on friction and impact in rigid-body
		dynamics~\cite{stewart00} at least if the normal reaction is considered
		to be given~\cite{stewart11}:
		\begin{equation}
			\contactforceCFto{j}(t) \in \argmin_{\norm{\vec y}_2 \leq \cof{j} \contactforceCFn{j}(t)} \vec y^\transp \postrelvelCFto{j}(t),
			\label{eq:maxdis}
		\end{equation}
		where the objective function corresponds to the (negated) rate of energy
		dissipation. Since the relative contact velocity at time~$t$ is
		independent of the contact reaction at time~$t$, the objective function
		is a \emph{linear} function of the frictional contact reaction at a
		non-impulsive point in time~$t \not \in \mathcal{T}_q$.

		When formulating the friction constraint on the impulsive contact
		reactions, the situation changes subtly but drastically: The impulsive
		contact reactions now influence the post-impulse relative contact
		velocity. The drastic consequence of this is that the maximum
		dissipation principle and the Coulomb friction model for that matter
		as it is formulated for non-impulsive contact reactions in
		Eq.~\eqref{eq:maxdis} cannot be transferred to impulsive contact
		reactions without in-depth modifications if the property of maximizing the
		energy dissipation is to be preserved. This statement stands in
		contrast to common practice~\cite{jean99, stewart00, bonnefon11}.
		In particular the
		term $\contactimpulseCFto{j}(t)^\transp \postrelvelCFto{j}(t)$ is a
		quadratic function of the impulsive contact reactions and it does not
		reflect the energy dissipated. Hence, impulsive frictional reactions
		directly opposing the relative contact velocity in the tangential plane
		also do not dissipate as much energy as allowed by the friction cone
		condition in general.

		The system energy~$E^{\sfrac{-}{+}}(t)$ is the sum of the potential
		energy~$U(t)$ and the kinetic energy $T^{\sfrac{-}{+}}(t)$:
		\begin{equation*}
			\begin{split}
				E^{\sfrac{-}{+}}(t) & = U(t) + T^{\sfrac{-}{+}}(t)\\
					& = U(t) + \frac{1}{2} \dvect{\linvel{}^{\sfrac{-}{+}}(t)}{\angvel{}^{\sfrac{-}{+}}(t)}^\transp \mat{M}(\orient{}(t)) \dvect{\linvel{}^{\sfrac{-}{+}}(t)}{\angvel{}^{\sfrac{-}{+}}(t)},
			\end{split}
		\end{equation*}
		where the potential energy is not affected by impulses. Insertion
		leads to the expression for the post-impulse system energy
		\begin{equation*}
			\shrinkeqnnew{1}{
			\begin{split}
				E^+(t) & = E^-(t) + \frac{1}{2} \contactimpulse{}(t)^\transp \mat{W}(t)^\transp \mat{M}(\orient{}(t))^{-1} \mat{W}(t) \contactimpulse{}(t) \\
				       & \hspace{-0.75cm} + \contactimpulse{}(t)^\transp \mat{W}(t)^\transp \left( \dvect{\linvel{}^-(t)}{\angvel{}^-(t)} + \mat{M}(\orient{}(t))^{-1} \dvect{\linimp{ext}(t)}{\angimp{ext}(t)} \right) \\
				       & \hspace{-0.75cm} + \dvect{\linimp{ext}(t)}{\angimp{ext}(t)}^\transp \left( \dvect{\linvel{}^-(t)}{\angvel{}^-(t)} + \frac{1}{2} \mat{M}(\orient{}(t))^{-1} \dvect{\linimp{ext}(t)}{\angimp{ext}(t)} \right) \\
				       & \hspace{-0.75cm} = \frac{1}{2} \contactimpulse{}(t)^\transp \mat{A}(t) \contactimpulse{}(t) - \contactimpulse{}(t)^\transp \vec b(t) + c_1(t)
			\end{split}
			}
		\end{equation*}
		in terms of the pre-impulse system energy and the impulsive contact
		reactions. Let $\mat{A}(t)$ be the Delassus	operator and let $\vec b(t)$
		condense the terms depending linearly on the impulsive contact reactions
		and let $c_1(t)$ condense the constant terms. A contact reaction
		$\contactimpulse{j}(t)$ complying with the maximum dissipation principle
		should minimize $E^+(t)$. Restating $E^+(t)$ in terms of the $j$-th
		contact reaction and assuming all other contact reactions to be constant results in
		\begin{equation*}
			\begin{split}
				E_j^+(\contactimpulse{j}(t)) & := \frac{1}{2} \contactimpulse{j}(t)^\transp \mat{A}_{jj}(t) \contactimpulse{j}(t) \\
				                             & - \contactimpulse{j}(t)^\transp \left( \vec b_j(t) - \mat{A}_{j\overline{j}}(t) \contactimpulse{\overline{j}}(t) \right) + c_2(t),
			\end{split}
		\end{equation*}
		where $\mat{A}_{jj}(t)$ corresponds to the $j$-th $3 \times 3$ diagonal
		block of the Delassus operator and where $\overline{j}$ selects all
		columns (elements) except for column~$j$ (element~$j$). The diagonal block
		can be determined to be~\cite{mirtich96,preclik14}
		\begin{equation}
			\shrinkeqnnew{0.9}{
			\begin{split}
				\mat{A}_{jj}(t) & = (\mass{a}^{-1} + \mass{b}^{-1}) \identmat{3} \\
				& - (\contactpos{j}(t) - \pos{a}(t))^\cross \inertia{aa}(\orient{a}(t))^{-1} (\contactpos{j}(t) - \pos{a}(t))^\cross \\
				& - (\contactpos{j}(t) - \pos{b}(t))^\cross \inertia{bb}(\orient{b}(t))^{-1} (\contactpos{j}(t) - \pos{b}(t))^\cross,
			\end{split}
			}
			\label{eq:diagblock}
		\end{equation}
		where $a = i_1(j)$ and $b = i_2(j)$.

		Then the impulsive contact reaction complying with the maximum dissipation principle is
		\begin{samepage}
			\begin{equation}
				\contactimpulse{j}(t) \in \argmin_{\substack{
						\tikz[baseline]{\node[font=\scriptsize,anchor=base,draw=diagfg,fill=diagbg,thick,rounded corners=2mm      ] (t1)  {$\norm{\vec \Lambda_{to}}_2 \leq \cof{j} \Lambda_n$};} \\
						\tikz[baseline]{\node[font=\scriptsize,anchor=base,draw=offdiagfg,fill=offdiagbg,thick,rounded corners=2mm] (t2)  {$\xi_j(t) \geq 0 \compl \Lambda_n \geq 0$};} \\
						\tikz[baseline]{\node[font=\scriptsize,anchor=base,draw=regfg,fill=regbg,thick,rounded corners=2mm        ] (t3)  {$\dot{\xi}_j^+(t) \geq 0 \compl \Lambda_n \geq 0$};} \\
						\tikz[baseline]{\node[font=\scriptsize,anchor=base                                                        ] (t4)  {$\vec n_j(t)^\transp \mat{A}_{jj}(t) \vec \Lambda \geq 0$};}
					}} E_j^+(\vec \Lambda),
				\label{eq:impmaxdis}
			\end{equation}
			\begin{tikzpicture}[overlay]
				\path[->,font=\scriptsize,line width=1pt] (t2) edge [out=180, in=180] node [left] {$\xi_j(t) = 0$} (t3);
				\path[anchor=east] let \p1 = (t1) in node at (\linewidth,\y1) {\scriptsize \inlinesubeqno \label{eq:fcone}};
				\path[anchor=east] let \p1 = (t2) in node at (\linewidth,\y1) {\scriptsize \inlinesubeqno \label{eq:signorini}};
				\path[anchor=east] let \p1 = (t3) in node at (\linewidth,\y1) {\scriptsize \inlinesubeqno \label{eq:impactlaw}};
				\path[anchor=east] let \p1 = (t4) in node at (\linewidth,\y1) {\scriptsize \inlinesubeqno \label{eq:nopressure}};
			\end{tikzpicture}
		\end{samepage}
		where Eq.~\eqref{eq:fcone} corresponds to the friction cone condition,
		Eq.~\eqref{eq:signorini} corresponds to the Signorini condition,
		Eq.~\eqref{eq:impactlaw} corresponds to the purely inelastic restitution hypothesis,
		and Eq.~\eqref{eq:nopressure} is an additional constraint requiring
		that the contact reaction is not increasing the contact pressure.
		The last constraint guarantees uniqueness for a single contact. It
		excludes non-zero solutions if the contact opens by itself.	The
		objective function is a quadratic function of the contact reactions
		and it is strictly convex since $\mat{A}_{jj}(t)$ is SPD.

		For open contacts ($\xi_j(t) > 0$) the Signorini constraint and the
		friction cone constraint restrict the feasible set to the reaction
		$\contactimpulse{j}(t) = \vec 0$. The restitution hypothesis is disabled and the
		pressure constraint is fulfilled.

		If the contact is closed~($\xi_j(t) = 0$) the Signorini condition
		reduces to $\contactimpulseCFn{j}(t) \geq 0$ and the restitution hypothesis is
		enabled. In order to determine whether the restitution hypothesis is
		active~($\dot{\xi}_j^+(t) = 0$) or inactive~($\dot{\xi}_j^+(t) > 0$),
		the dependence of $\dot{\xi}_j^+(t)$ on the impulsive contact reactions
		$\contactimpulse{}(t)$ must become explicit. At least for spherical
		particles the time-derivative of the post-impact signed distance
		function can be expressed in terms of the relative contact
		velocities:
		\begin{equation}
			\begin{split}
				\dot{\xi}_j^+(t) & = \postrelvelCFn{j}(t)\\
				                 & \hspace{-0.5cm} = \vec n_j(t)^\transp (\mat{A}_{jj}(t) \contactimpulse{j}(t) + \mat{A}_{j\overline{j}}(t) \contactimpulse{\overline{j}}(t) - \vec b_j(t)).
			\end{split}
			\label{eq:maxcompr}
		\end{equation}

		Contacts fulfilling the property
		$\vec n_j(t)^\transp (\mat{A}_{j\overline{j}}(t) \contactimpulse{\overline{j}}(t) - \vec b_j(t)) < 0$,
		that is contacts where a penetration is imminent if no impulsive contact
		reaction acts, are termed \emph{colliding} in the following. Pre-impulse
		velocities, external impulses and impulsive reactions from other
		simultaneously colliding contacts determine if the contact is in a
		colliding state.

		Contacts fulfilling the property $\vec n_j(t)^\transp (\mat{A}_{j\overline{j}}(t) \contactimpulse{\overline{j}}(t) - \vec b_j(t)) > 0$,
		that is contacts where separation is imminent if no impulsive contact reaction acts, are
		termed \emph{separating}. For separating closed contacts, the reaction $\contactimpulse{j}(t) = \vec 0$
		fulfills all constraints and thus the restitution hypothesis is inactive. The
		pressure constraint ensures that no non-zero solutions exist.
		
		For colliding closed contacts, the restitution hypothesis must be active,
		restricting the feasible set to the plane of maximum compression
		defined by Eq.~\eqref{eq:maxcompr}. Combined with the friction cone
		condition, the feasible set forms a conic section. The normal of the
		plane of maximum compression is $\mat{A}_{jj}(t) \vec n_j(t)$. Since
		$\mat{A}_{jj}(t)$ is SPD and since the contact is colliding, the conic
		section is guaranteed to be non-empty. Since the conic sections are
		non-empty convex sets and since the objective function is strictly
		convex, the optimization problem has a unique global minimum. The
		pressure condition is fulfilled since it is fulfilled for any point on
		the plane of maximum compression in the colliding case.
		
		The unconstrained global minimum $\vec \Lambda_0$ is given by
		\begin{equation*}
			\nabla E_j^+ (\vec \Lambda_0) = \mat{A}_{jj}(t) \vec \Lambda_0 - (\vec b_j(t) - \mat{A}_{j\overline{j}}(t) \contactimpulse{\overline{j}}(t))= \vec 0.
		\end{equation*}
		If the contact is colliding and closed and if $\vec \Lambda_0$ fulfills
		all constraints, then $\contactimpulse{j}(t) = \vec \Lambda_0$ and the
		post-impulse relative contact velocity in the tangential plane is zero
		corresponding to a sticking~(static) post-impulse contact state.
		If $\vec \Lambda_0$ does not fulfill all constraints the post-impulse
		relative contact velocity in the tangential plane is non-zero and the
		post-impulse contact state thus sliding~(dynamic). The
		solution~$\contactimpulse{j}(t)$ of the contact problem then resides on
		the boundary of the conic section.

		The friction model for contact impulses from Eq.~\eqref{eq:impmaxdis}
		thus maximizes dissipation by minimizing the (kinetic) energy
		analogously to the velocity-level constraint of the Coulomb model for
		non-impulsive contact reactions. This is in contrast to the alleged
		extension of the Coulomb model to impulsive reactions, where the
		dissipation is not (sufficiently) maximized. At the same time the
		contact impulses fulfill the friction cone condition as in the Coulomb
		model and they fulfill the inelastic restitution hypothesis. The friction model
		is lazy in the sense that separating solutions (no contact reaction)
		are preferred over any other solutions, which is guaranteed through the
		pressure condition. Furthermore, static solutions (no work performed)
		are preferred over dynamic solutions analogously to the
		acceleration-level constraint of the Coulomb model.

		\section{Numerical Methods}
		\label{sec:numerics}

		\subsection{Multi-Contact Problems}

			Integrating the equations describing the rigid-body dynamics can be
			approached in at least two different ways: The event-driven approach aims to
			predict the next impulsive point in time~$\tImp \in \mathcal{T}_q$
			given an initial state at time~$t_0$. Then the simulation is integrated
			until~$\tImp$ assuming the impulsive
			contact reactions~$\contactimpulse{}(t)$ to be zero for
			$t \in \intervalOO{t_0}{\tImp}$. At time~$\tImp$ an impact
			problem given by
			\begin{equation*}
				\dvect{\linvel{}^{+}(\tImp)}{\angvel{}^{+}(\tImp)} = \dvect{\linvel{}^-(\tImp)}{\angvel{}^-(\tImp)} + \mat{M}(\orient{}(\tImp))^{-1} \dvect{ \linimp{}(\tImp)}{\angimp{}(\tImp)}
			\end{equation*}
			and Eq.~\eqref{eq:impmaxdis} has to be solved. Then, the integration can
			be restarted having a state fulfilling all constraints at hand.
			The difficulty of this
			approach lies in the problem of predicting the next impulsive point
			in time, which is a priori unknown. If rigid bodies follow ballistic
			trajectories impact times can be predicted accurately and efficient
			simulation codes exist~\cite{mirtich96, bannerman11}. For the more
			general case, where e.g.~impulsive points in time can not only stem
			from collisions but also stem from self-locking sliding frictional
			contacts~\cite{shen11}, we know of no efficient method to
			accurately predict the next impulsive point in time. Furthermore,
			simulation codes necessarily stall in situations, where the collision
			frequency increases unboundedly like in cases where a bouncing ball
			comes to rest on a plane.

			The second category of approaches for integrating rigid-body dynamics
			are the time-stepping methods. These methods proceed in time
			steps~$\dt > 0$ independent of the impulsive points in time.
			The methods do not distinguish between non-impulsive and
			impulsive contact reactions but implicitly solve for integrals of
			the contact reactions~$\contactforceapprox{}$. The integrals of the contact reactions are
			then constrained to fulfill contact conditions at selected points
			in time. In the following an impulse-velocity time-stepping
			scheme is used, which is described in detail in~\cite{preclik14}. It is similar to the
			schemes by Anitescu~\cite{anitescu97}, Tasora~\cite{tasora11} and
			Stewart~\cite{stewart96}. The equations of motion are integrated
			using a discretization similar to a semi-implicit Euler method,
			where positions are integrated implicitly and velocities are
			integrated explicitly:
			\begin{equation*}
			\shrinkeqnnew{1}{
				\begin{split}
					\pos{}'(\contactforceapprox{})    & = \pos{}    + \dt \linvel{}'(\contactforceapprox{}), \\
					\orient{}'(\contactforceapprox{}) & = \sfrac{\left(\orient{} + \dt \mat{Q}(\orient{}) \angvel{}'(\contactforceapprox{})\right)}{\norm{\orient{} + \dt \mat{Q}(\orient{}) \angvel{}'(\contactforceapprox{})}_2}, \\
					\dvect{\linvel{}'(\contactforceapprox{})}{\angvel{}'(\contactforceapprox{})} & = \dvect{\linvel{}}{\angvel{}} + \mat{M}(\orient{})^{-1} \left( \mat{W} \contactforceapprox{} + \dvect{\forceextapprox{}}{\torqueextapprox{} - \dt \angvel{} \cross \inertia{}(\orient{}) \angvel{}} \right),
				\end{split}
			}
			\end{equation*}
			where primes identify state variables at time $t + \dt$ in contrast to
			state variables	at time~$t$ omitting the prime. The relative contact
			velocities are
			\begin{equation*}
				\primerelvel{}(\contactforceapprox{}) = \mat{W}^\transp \dvect{\linvel{}'(\contactforceapprox{})}{\angvel{}'(\contactforceapprox{})}.
			\end{equation*}
			The discrete non-penetration constraint for a contact~$j$ is
			\begin{equation*}
				\frac{\xi_j}{\dt} + \primerelvelCFn{j}(\contactforceapprox{}) \geq 0 \compl \contactforceapproxCFn{j} \geq 0,
			\end{equation*}
			which if the gap is closed exactly ($\xi_j = 0$) states that the relative contact velocity at the end of the time step and in the
			direction of the contact normal must be non-approaching and complementary to the
			non-adhesive contact reaction in normal direction. The term $\frac{\xi_j}{\dt}$ acts as an error reduction term
			if penetrations are present~($\xi_j < 0$). In that case it can be scaled
			down when needed using an error reduction parameter $\varepsilon \in \intervalCO{0}{1}$
			to avoid introducing an excessive amount of energy ($\varepsilon \min(0, \frac{\xi_j}{\dt}) + \max(0, \frac{\xi_j}{\dt})$).
			If a positive gap is present~($\xi_j > 0$), the term ensures that contact reaction remains zero if the contact would not close within the time step.
			The friction cone condition
			\begin{equation*}
				\norm{\contactforceapproxCFto{j}}_2 \leq \cof{j} \contactforceapproxCFn{j}
			\end{equation*}
			is adopted as it stands. Instead of requiring the impulsive contact
			reactions to fulfill the conventional Coulomb friction constraints
			as in
			\begin{equation*}
				\norm{\primerelvelCFto{j}(\contactforceapprox{})}_2 \contactforceapproxCFto{j} = -\cof{j} \contactforceapproxCFn{j} \primerelvelCFto{j}(\contactforceapprox{}),
			\end{equation*}
			the maximum dissipation principle from Eq.~\eqref{eq:impmaxdis} is used, since
			the discretized relative contact velocity in the tangential plane depends on the contact
			reactions as before. Therefore, the energy term is discretized, leading to
			\begin{equation*}
				\begin{split}
					E'(\contactforceapprox{}) & = U + T'(\contactforceapprox{}) = \frac{1}{2} \contactforceapprox{}^\transp \mat{W}^\transp \mat{M}(\orient{})^{-1} \mat{W} \contactforceapprox{} \\
						& \hspace{-0.75cm} + \contactforceapprox{}^\transp \mat{W}^\transp \left( \dvect{\linvel{}}{\angvel{}} + \mat{M}(\orient{})^{-1} \dvect{\forceextapprox{}}{\torqueextapprox{} - \dt \angvel{} \cross \inertia{}(\orient{}) \angvel{}} \right) \\
						& \hspace{-0.75cm} + \vec z^\transp \left( \dvect{\linvel{}}{\angvel{}} + \frac{1}{2} \mat{M}(\orient{})^{-1} \vec z \right) + E \\
						& \hspace{-0.75cm} = \frac{1}{2} \contactforceapprox{}^\transp \mat{A} \contactforceapprox{} - \contactforceapprox{}^\transp \vec b + c_1,
				\end{split}
			\end{equation*}
			where $\vec z = \left( \forceextapprox{}^\transp, (\torqueextapprox{} - \dt \angvel{} \cross \inertia{}(\orient{}) \angvel{})^\transp \right)^\transp$.
			For a single contact~$j$ the energy term reduces to
			\begin{equation*}
				E_j'(\contactforceapprox{j}) := \frac{1}{2} \contactforceapprox{j}^\transp \mat{A}_{jj} \contactforceapprox{j} - \contactforceapprox{j}^\transp (\vec b_j - \mat{A}_{j\overline{j}} \contactforceapprox{\overline{j}}) + c_2
			\end{equation*}
			given all other contact reactions~$\contactforceapprox{\overline{j}}$. The relative contact velocity in
			terms of $\contactforceapprox{j}$ and $\contactforceapprox{\overline{j}}$ is
			\begin{equation*}
				\primerelvel{j}(\contactforceapprox{j}) := \mat{A}_{jj} \contactforceapprox{j} - (\vec b_j - \mat{A}_{j\overline{j}} \contactforceapprox{\overline{j}}).
			\end{equation*}
			Hence, the contact constraint complying with the maximum dissipation principle is
			\begin{equation}
				\contactforceapprox{j} \in \argmin_{\substack{
						\norm{\contactforceCFto{}}_2 \leq \cof{j} \contactforceCFn{} \\
						\frac{\xi_j}{\dt} + \primerelvelCFn{j}(\contactforce{}) \geq 0 \compl \contactforceCFn{} \geq 0 \\
						\vec n_j^\transp \mat{A}_{jj} \contactforce{} \geq 0
					}} E_j'(\contactforce{}).
				\label{eq:discretemaxdis}
			\end{equation}
			Then a solution $\contactforceapprox{}$ of the multi-contact time-step problem satisfies Eq.~\eqref{eq:discretemaxdis} for all contacts~$j = 1\twodots\numcontacts$.
			The problem can be solved using a non-linear block Gauss-Seidel or variants thereof as demonstrated in~\cite{preclik14,preclik15}.
			The solution algorithm proposed there reduces the multi-contact problem to the problem of solving a sequence of single-contact problems.
			Hence, in the next section an analytic solution for the single-contact impact problem from Eq.~\eqref{eq:discretemaxdis} (and Eq.~\eqref{eq:impmaxdis} alike) is derived.

		\subsection{Single-Contact Problems}

			The single-contact problem for impulsive contact reactions complying with the maximum dissipation principle from
			Eq.~\eqref{eq:impmaxdis} constrains solutions to zero if contacts are open~($\xi_j(t) > 0$). If contacts are closed,
			Eq.~\eqref{eq:impmaxdis} has the same structure as the discrete single-contact time-step problem from
			Eq.~\eqref{eq:discretemaxdis}:
			\begin{equation*}
				\vec x_* \in \argmin_{\substack{\norm{\vec x_{to}}_2 \leq \cof{} x_n \\ \vec{a}_n^\transp \vec x - b_n \geq 0 \compl x_n \geq 0 \\ \vec a_n^\transp \vec x \geq 0}} \frac{1}{2} \vec x^\transp \mat{A} \vec x - \vec x^\transp \vec b,
			\end{equation*}
			where $\cof{} \in \R^{\geq 0}$, $\mat{A}$ is SPD and
			\begin{equation*}
				\begin{split}
					\mat{A} & = \begin{bmatrix} A_{nn} & A_{nt} & A_{no} \\ A_{nt} & A_{tt} & A_{to} \\ A_{no} & A_{to} & A_{oo} \end{bmatrix} \in \R^{3 \times 3}, \vec a_n = \begin{pmatrix} A_{nn} \\ A_{nt} \\ A_{no} \end{pmatrix} \in \R^3, \\
					\vec x & = \begin{pmatrix} x_n \\ x_t \\ x_o \end{pmatrix} \in \R^3, \vec x_{to} = \begin{pmatrix} x_t \\ x_o \end{pmatrix} \in \R^2, \vec b = \begin{pmatrix} b_n \\ b_t \\ b_o \end{pmatrix} \in \R^3.
				\end{split}
			\end{equation*}
			If the contact is separating~($b_n < 0$) the solution is constrained to $\vec x_* = \vec 0$ since the
			pressure condition~($\vec a_n^\transp \vec x \geq 0$) prevents any non-zero solutions.
			If the contact is non-separating~($b_n \geq 0$) the pressure condition is redundant and the constraint
			set is formed by the intersection of the plane of maximum compression and the friction cone.
			This conic section is non-empty but can take on the shape of an ellipse, parabola or hyperbola.
			The cases where $b_n = 0$ are special in the sense that the plane of maximum compression includes
			the origin and thus the conic section degenerates to a point, a ray or a degenerate hyperbola.
			These cases will be discussed after the cases where $b_n > 0$.
			The cases where $b_n > 0$ and $\cof{} = 0$ are also special
			since the conic section degenerates to a single (non-zero) point $\vec x_* = (A_{nn}^{-1} b_n, 0, 0)^\transp$.
			If $b_n > 0$ and $\cof{} > 0$ the conic section is non-degenerate. The unconstrained minimum of
			the objective function is $\vec x_0 = \mat{A}^{-1} \vec b$. If it fulfills the friction cone condition it is
			contained in the constraint set and it thus solves the contact problem $\vec x_* = \vec x_0$.
			
			If the unconstrained minimum of the objective function does not fulfill the friction cone condition,
			the solution must be located on the boundary of the conic section minimizing the objective function.
			Eliminating the normal component using the equation for the plane of
			maximum compression and switching to polar coordinates leads to
			\begin{equation*}
				\shrinkeqnnew{1}{
				\begin{split}
					\vec x & = \left(A_{nn}^{-1}(b_n - A_{nt} x_t - A_{no} x_o), x_t, x_o\right)^\transp \\
					       & = \left(A_{nn}^{-1}(b_n - A_{nt} r \cos \alpha - A_{no} r \sin \alpha), r \cos \alpha, r \sin \alpha\right)^\transp,
				\end{split}
				}
			\end{equation*}
			where $r \in \R^{\geq 0}$ and $\alpha \in \R$. The friction cone
			condition then becomes
			\begin{equation*}
				\shrinkeqnnew{1}{
				\begin{split}
					                          & \norm{\vec x_{to}}_2 \leq \cof{} A_{nn}^{-1}(b_n - A_{nt} x_t - A_{no} x_o) \\
					\leftrightarrow \quad & r (A_{nn} + \cof{} A_{nt} \cos \alpha + \cof{} A_{no} \sin \alpha) \leq \cof{} b_n \\
					\leftrightarrow \quad & r \underbrace{(A_{nn} + \cof{} \sqrt{A_{nt}^2 + A_{no}^2} \cos (\alpha - \atan2(A_{no}, A_{nt})))}_{=f_r(\alpha)} \leq \cof{} b_n.
				\end{split}
				}
			\end{equation*}
			For angles~$\alpha$ where $f_r(\alpha) \leq 0$, the inequality poses
			no additional restrictions on the non-negative coordinate~$r$. For
			angles~$\alpha$ where $f_r(\alpha) > 0$, the inequality defines an
			upper bound on the coordinates~$r$. In both cases the component~$r$
			satisfies
			\begin{equation*}
				0 \leq r \leq \frac{\cof{} b_n}{\max(0, f_r(\alpha))}.
			\end{equation*}
			Let $\mathcal{I} + 2\pi \mathbb{N}$ be the set of angles for which
			$f_r(\alpha) > 0$ holds. $\mathcal{I}$ can be determined to be
			\begin{equation}
				\mathcal{I} = \begin{cases}
					\intervalOO{-\Delta \alpha}{\Delta \alpha} + \alpha_0 & \text{if $A_{nn} \leq \cof{} \sqrt{A_{nt}^2 + A_{no}^2}$} \\
					\intervalCO{0}{2\pi} & \text{else}
				\end{cases},
				\label{eq:feasible}
			\end{equation}
			where $\Delta \alpha = \acos \frac{-A_{nn}}{\cof{}\sqrt{A_{nt}^2 + A_{no}^2}}$ and $\alpha_0 = \atan2(A_{no}, A_{nt})$. Let
			\begin{equation*}
				\vec \gamma(\alpha) = \overline{r}(\alpha)\dvect{\cos\alpha}{\sin\alpha}
			\end{equation*}
			describe the curve along the boundary of the conic section, where $\overline{r}(\alpha) = \frac{\mu b_n}{f_r(\alpha)}$.
			Then the contact problem reduces to
			\begin{equation*}
				\begin{split}
					\alpha_* & \in 2\pi \mathbb{N} + \argmin_{\alpha \in \mathcal{I}} \overbrace{\frac{1}{2} \vec \gamma(\alpha)^\transp \mat{\hat{A}} \vec \gamma(\alpha) - \vec \gamma(\alpha)^\transp \vec{\hat{b}}}^{=f_{obj}(\alpha)} \\
					r_*      & = \overline{r}(\alpha_*),
				\end{split}
			\end{equation*}
			where
			\begin{equation*}
				\begin{split}
					\mat{\hat{A}} & = \begin{bmatrix} A_{tt} - A_{nn}^{-1} A_{nt}^2 & A_{to} - A_{nn}^{-1} A_{nt} A_{no} \\ A_{to} - A_{nn}^{-1} A_{nt} A_{no} & A_{oo} - A_{nn}^{-1} A_{no}^2 \end{bmatrix}\ \text{and} \\
					\vec{\hat{b}} & = \dvect{b_t - A_{nn}^{-1} A_{nt} b_n}{b_o - A_{nn}^{-1} A_{no} b_n}
				\end{split}
			\end{equation*}
			result from eliminating the normal component. The objective
			function~$f_{obj}$ is univariate and $2\pi$-periodic but no longer
			strictly convex nor quadratic.
			\begin{figure}
				\centering
				\floatbox{figure}[\textwidth]
					{
						\begin{subfloatrow}[1]%
							\floatbox{figure}
								{\includegraphics[width=0.7\FBwidth]{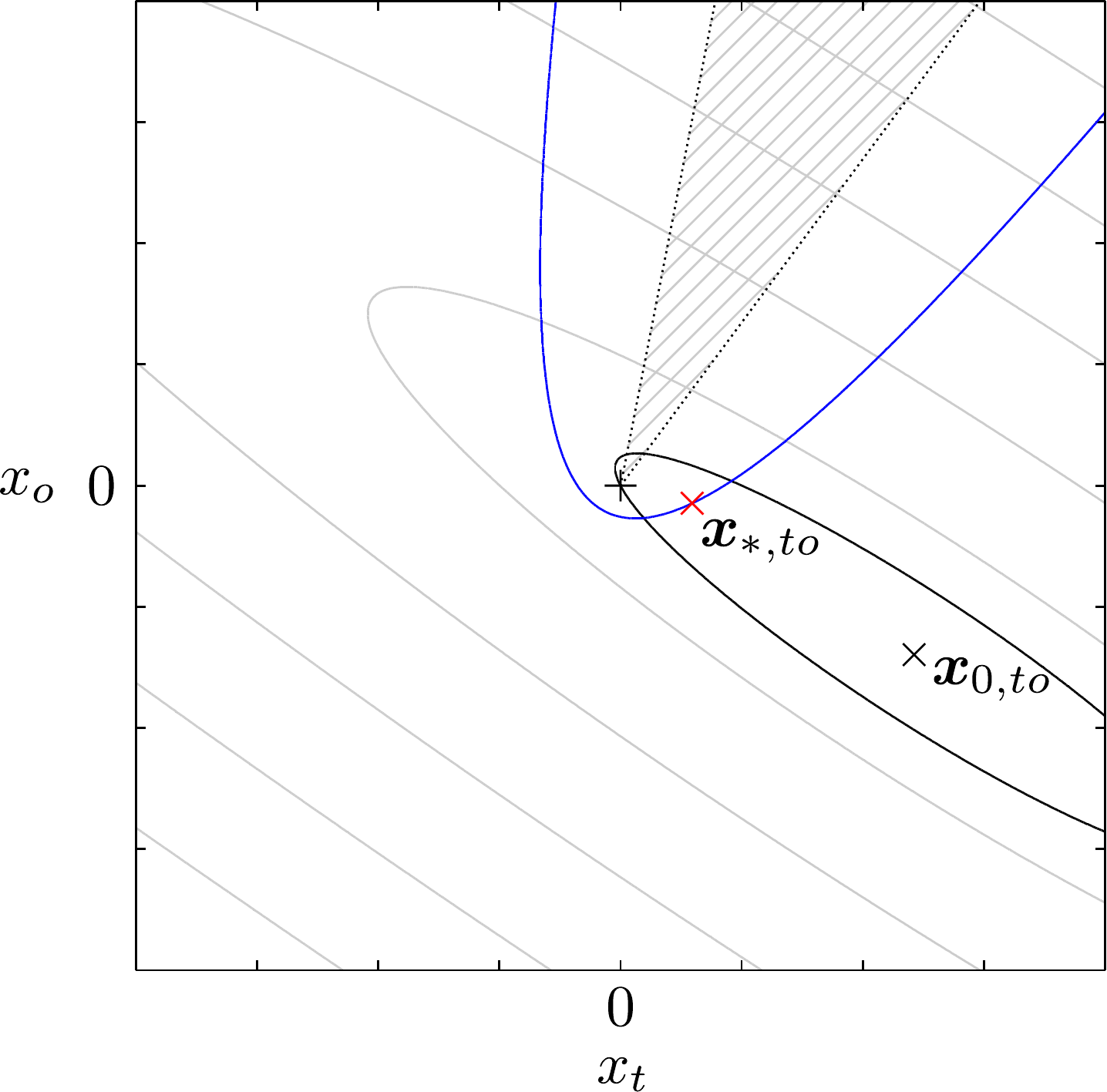}}
								{\caption{The ellipsoidal contour lines of $\frac{1}{2} \vec x_{to}^\transp \mat{\hat{A}} \vec x_{to} - \vec x_{to}^\transp \vec{\hat{b}}$ are solid gray except the zero contour line which is solid black. The conic section is a hyperbola and drawn as a solid blue curve. The unconstrained minimum is marked by a black cross. The red cross is the maximally dissipative solution along the hyperbola.}\label{fig:hyperbola1a}}
						\end{subfloatrow}
						\vspace{\parskip}

						\begin{subfloatrow}[1]%
							\floatbox{figure}
								{\caption{The objective function $f_{obj}$ is plotted in solid blue. The non-feasible region is indicated by gray stripes. The poles are marked by dotted lines. The global minimum is marked by a red cross.}\label{fig:hyperbola1b}}
								{\includegraphics[width=\FBwidth]{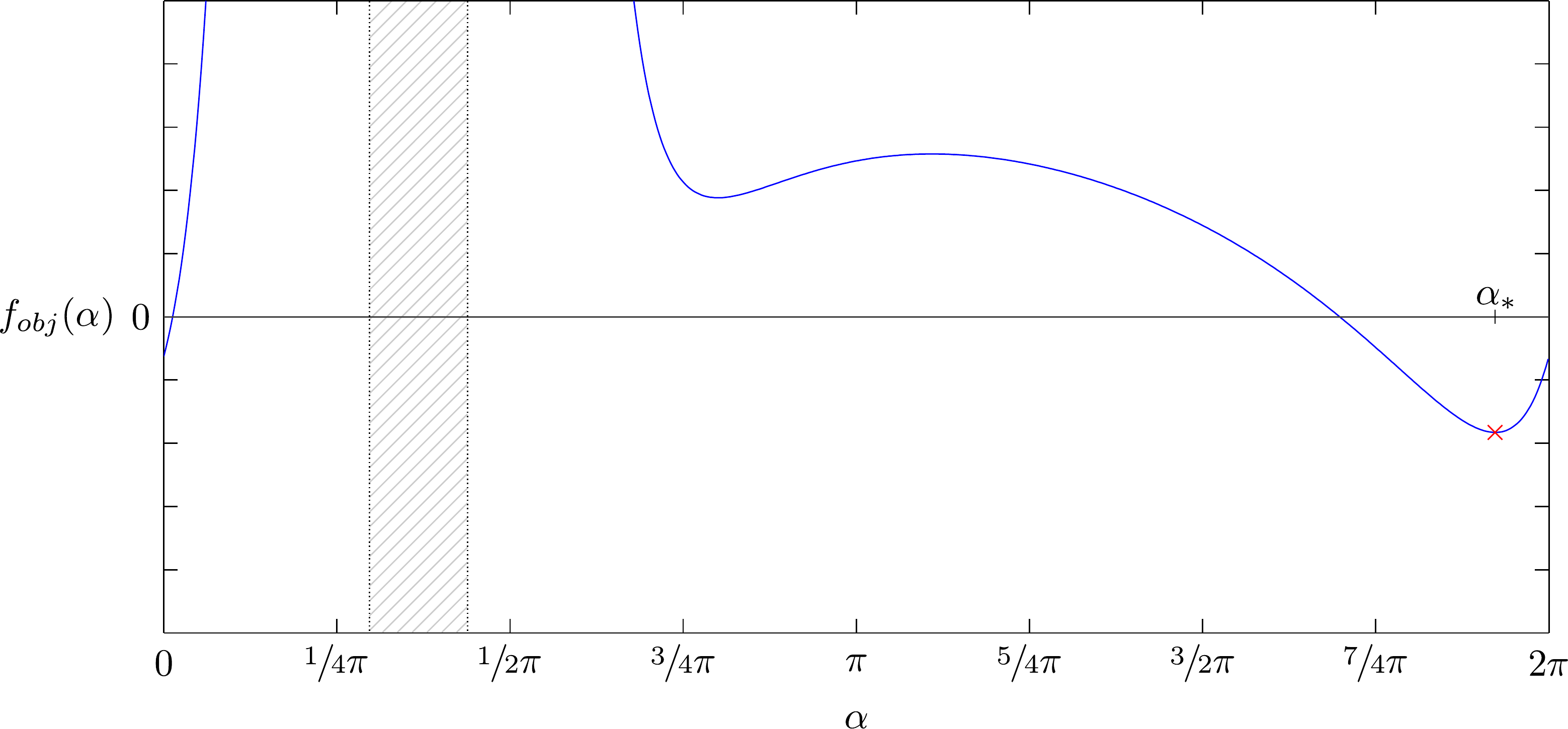}}
						\end{subfloatrow}
					}
					{\caption{An exemplary contact problem, where $\mat{A}$, $\vec{b}$ and $\cof{}$ are such that the conic section is a hyperbola and the objective function $f_{obj}$ has two local minima, but only one global minimum. Only the global minimum is dissipative.}\label{fig:hyperbola1}}
			\end{figure}
			\figref{fig:hyperbola1} illustrates the optimization problem for an
			exemplary (dynamic) contact. \figref{fig:hyperbola1a} plots the
			quadratic objective function for points~$\vec x_{to}$ on the plane
			of maximum compression and overlays the conic section.
			\figref{fig:hyperbola1b} plots the corresponding non-linear
			objective function~$f_{obj}$ in comparison. An iterative approach
			for solving this constrained minimization problem that is
			guaranteed to converge linearly is derived in~\cite{preclik14}.
			In the following an analytic approach is presented.

			Let
			\begin{equation*}
				\shrinkeqnnew{1}{
				\vec t(\alpha)
	= \dvect{\od{}{\beta}\overline{r}(\beta)\sin\beta\sVert[1]_{\beta=\alpha}}{\od{}{\beta}\overline{r}(\beta)\cos\beta\sVert[1]_{\beta=\alpha}}
	= \dvect{\od{\overline{r}(\beta)}{\beta}\sVert[1]_{\beta=\alpha} \cos \alpha - \overline{r}(\alpha) \sin \alpha}{\od{\overline{r}(\beta)}{\beta}\sVert[1]_{\beta=\alpha} \sin \alpha + \overline{r}(\alpha) \cos \alpha}
				}
			\end{equation*}
			be the unit vector tangential to the curve, where
			\begin{equation*}
				\shrinkeqnnew{1}{
				\od{\overline{r}(\beta)}{\beta}\sVert[1]_{\beta=\alpha} = \frac{\overline{r}(\alpha)^2}{b_n} \sqrt{A_{nt}^2 + A_{no}^2} \sin(\alpha - \atan2(A_{no}, A_{nt})).
				}
			\end{equation*}
			Angles minimizing $f_{obj}$ must satisfy
			\begin{equation}
				\vec t(\alpha)^\transp (\mat{\hat{A}} \vec \gamma(\alpha) - \vec{\hat{b}}) = 0.
				\label{eq:tangent}
			\end{equation}
			Insertion, trigonometric identity transformations and
			multiplication by $\frac{\mu^2 b_n^2}{\overline{r}(\alpha)^3}$ leads to
			a trigonometric equation in the form of
			\begin{equation}
				c_0 + c_1 \cos(\alpha + \varphi_1) + c_2 \cos(2\alpha + \varphi_2) = 0,
				\label{eq:trig}
			\end{equation}
			with constants $c_0, c_1, c_2, \varphi_1, \varphi_2 \in \R$ as
			specified in appendix~\ref{sec:appendix}. The trigonometric equation
			can be transformed into a quartic equation by substituting
			$y = e^{i \alpha} = \cos \alpha + i \sin \alpha$. After solving
			the quartic equation for $y_i$, the corresponding angles
			$\alpha_i = \atan2(\operatorname{Im}(y_i), \operatorname{Re}(y_i))$ have to be checked for
			validity. Angles $\alpha_i \not \in 2\pi\mathbb{N}+\mathcal{I}$ are invalid as well as angles not satisfying Eq.~\eqref{eq:tangent} or Eq.~\eqref{eq:trig} for that matter. Among
			all other candidates, the one with minimum objective function value
			amounts to the $2\pi$-periodic solution $\alpha_*$. Finally,
			\begin{equation*}
				\vec x_* = \dvect{A_{nn}^{-1}(b_n - A_{nt} \overline{r}(\alpha_*) \cos \alpha_* - A_{no} \overline{r}(\alpha_*) \sin \alpha_*)}{\vec \gamma(\alpha_*)}.
			\end{equation*}

			Back to the case, where $b_n = 0$ and $\cof{} > 0$: If the conic section corresponds to
			a degenerate ellipse ($A_{nn} > \cof{} \sqrt{A_{nt}^2 + A_{no}^2}$),
			then the solution is exactly this point. The conic sections that
			correspond to a degenerate parabola ($A_{nn} = \cof{} \sqrt{A_{nt}^2 + A_{no}^2}$)
			are equivalent to the ray
			\begin{equation*}
				\setprop{r\dvect{A_{nt}}{A_{no}}}{r \in \R^{\geq 0}}.
			\end{equation*}
			Degenerate hyperbolas correspond to the conical combination of two rays (along the asymptotes)
			\begin{equation*}
				\shrinkeqnnew{1}{
				\setprop{r_1\dvect{\cos (\alpha_0 - \Delta \alpha)}{\sin (\alpha_0 - \Delta \alpha)} + r_2 \dvect{\cos (\alpha_0 + \Delta \alpha)}{\sin (\alpha_0 + \Delta \alpha)}}{r_1, r_2 \in \R^{\geq 0}},
				}
			\end{equation*}
			where $\alpha_0$ and $\Delta \alpha$ as in Eq.~\eqref{eq:feasible}.
			Thus degenerate parabolas are a limiting case of degenerate hyperbolas with $\Delta \alpha = \pi$.
			Minimizing the energy over degenerate parabolas and hyperbolas can
			be implemented by checking, whether the unconstrained minimum $\vec x_0$ is contained
			in the friction cone. If that is not the case, then the objective function
			needs to be minimized along one of the rays:
			\begin{equation}
				\shrinkeqnnew{0.9}{
				\begin{split}
					\alpha_* & \in 2\pi\mathbb{N} + \begin{cases}
						\alpha_0 - \Delta \alpha & \text{if $(A_{no}, -A_{nt})^\transp \vec x_{0,to} \geq 0$} \\
						\alpha_0 + \Delta \alpha & \text{else}
					\end{cases}, \\
					r_* & = \max(0, (\dvect{\cos\alpha_*}{\sin\alpha_*}^\transp \mat{\hat{A}} \dvect{\cos\alpha_*}{\sin\alpha_*})^{-1} \dvect{\cos\alpha_*}{\sin\alpha_*}^\transp \vec{\hat{b}}).
				\end{split}
				}
				\label{eq:degenerate}
			\end{equation}
			The structogram in~\figref{fig:structogram} summarizes the analytic
			solutions of the various cases involved in solving the single-contact
			problem.

			\begin{figure*}
				\resizebox{\textwidth}{!}{
					\begin{tabular}{@{}c@{}}
						\begin{tikzpicture}
							\matrix (sg) [table,ampersand replacement=\&] {
								\& \& \& \& \& \& \\
								\& \& \& \& \& \& \\
								\& \& \& \& \& \& \\
								\& \& \& \& \& \& \\
								\& \& \& \& \& \& \\
								\& \& \& \& \& \& \\
							};

							\draw[black] (sg-1-1.north west) -- (sg-1-7.north east);
							\draw[black] (sg-2-2.north west) -- (sg-2-7.north east);
							\draw[black] (sg-3-3.north west) -- (sg-3-7.north east);
							\draw[black] (sg-4-4.north west) -- (sg-4-7.north east);
							\draw[black] (sg-5-5.north east) -- (sg-5-7.north east);
							\draw[black] (sg-6-5.north west) -- (sg-6-7.north west);
							\draw[black] (sg-6-1.south west) -- (sg-6-7.south east);

							\draw[black] (sg-1-1.north west) -- (sg-6-1.south west);
							\draw[black] (sg-1-2.north west) -- (sg-6-2.south west);
							\draw[black] (sg-2-3.north west) -- (sg-6-3.south west);
							\draw[black] (sg-3-4.north west) -- (sg-6-4.south west);
							\draw[black] (sg-4-5.north west) -- (sg-6-5.south west);
							\draw[black] (sg-4-6.north west) -- (sg-5-6.south west);
							\draw[black] (sg-5-7.north west) -- (sg-6-7.south west);
							\draw[black] (sg-1-7.north east) -- (sg-6-7.south east);

							\draw[black] (sg-1-1.north west) -- (sg-1-1.south east);
							\draw[black] (sg-2-2.north west) -- (sg-2-2.south east);
							\draw[black] (sg-3-3.north west) -- (sg-3-3.south east);
							\draw[black] (sg-4-4.north west) -- (sg-5-4.south east);
							\draw[black] (sg-4-5.north west) -- (sg-4-5.south east);

							\draw[black] (sg-6-1.south west) -- (sg-2-1.north east);
							\draw[black] (sg-6-2.south west) -- (sg-3-2.north east);
							\draw[black] (sg-6-3.south west) -- (sg-4-3.north east);
							\draw[black] (sg-6-4.south west) -- (sg-6-4.north east);
							\draw[black] (sg-5-5.south west) -- (sg-5-5.north east);

							\node[right, align=left]   at (sg-1-1.east)                      {$\vec x_* = \vec 0$};
							\node[right, align=left]   at (sg-2-2.east)                      {$\vec x_* = (A_{nn}^{-1}b_n, 0, 0)^\transp$};
							\node[right, align=left]   at (sg-3-3.east)                      {$\vec x_* = \vec x_0$};
							\node[right, align=left]   at (sg-4-5.east)                      {$\vec x_* = \vec 0$};
							\node[right, align=left]   at (sg-5-5.east)                      {$\alpha_*, r_*$ from Eq.~\eqref{eq:degenerate}};
							\node[right, align=left]   at (sg-6-4.east)                      {\begin{varwidth}{5cm}Let $\mathcal{S}$ be the sol. set of Eq.~\eqref{eq:trig}. \\ $\displaystyle \alpha_* \in 2\pi \mathbb{N} + \argmin_{\alpha \in \mathcal{I} \cap \mathcal{S}} f_{obj}(\alpha)$ \\ $r_* = \overline{r}(\alpha_*)$ \end{varwidth}};
							\node[right, align=left]   at ([yshift=-2.5ex]sg-6-6.north east) {\begin{varwidth}{5cm} $x_{*,n} = A_{nn}^{-1}(b_n - A_{nt} r_* \cos \alpha_*$ \\ \hphantom{$x_{*,n} = A_{nn}^{-1}(b_n$} $ - A_{no} r_* \sin \alpha_*)$ \\ $\vec x_* = \tvect{x_{*,n}}{r_* \cos \alpha_*}{r_* \sin \alpha_*}$\end{varwidth}};

							\node[below left, align=right, inner sep=2pt] at (sg-1-1.north east) {\footnotesize true};
							\node[below left, align=right, inner sep=2pt] at (sg-2-2.north east) {\footnotesize true};
							\node[below left, align=right, inner sep=2pt] at (sg-3-3.north east) {\footnotesize true};
							\node[below left, align=right, inner sep=2pt] at (sg-4-4.north east) {\footnotesize true};
							\node[below left, align=right, inner sep=2pt] at (sg-4-5.north east) {\footnotesize true};
							
							\node[above left, align=right, inner sep=2pt] at (sg-6-1.south east) {\footnotesize false};
							\node[above left, align=right, inner sep=2pt] at (sg-6-2.south east) {\footnotesize false};
							\node[above left, align=right, inner sep=2pt] at (sg-6-3.south east) {\footnotesize false};
							\node[above left, align=right, inner sep=2pt] at (sg-6-4.south east) {\footnotesize false};
							\node[above left, align=right, inner sep=2pt] at (sg-5-5.south east) {\footnotesize false};

							\node[yshift=-1ex, font=\small] at (sg-1-1.south) {$b_n < 0$};
							\node[yshift=-1ex, font=\small] at (sg-2-2.south) {$\cof{} = 0$};
							\node[yshift=-1ex, font=\small] at (sg-3-3.south) {$P_1$};
							\node[font=\small] at (sg-5-4.south) {$b_n = 0$};
							\node[font=\small] at (sg-4-5.south) {$P_2$};
						\end{tikzpicture}\\
						$P_1 = (x_{0,n} \geq 0 \land \norm{\vec x_{0,to}}_2 \leq \cof{} x_{0,n}) \qquad P_2 = (A_{nn} > \cof{} \sqrt{A_{nt}^2 + A_{no}^2})$
					\end{tabular}
				}

				\caption{The structogram describing the analytic solution of the single-contact problem.\label{fig:structogram}}
			\end{figure*}

	\section{Numerical Results}
	\label{sec:results}

	\subsection{Academic Single-Contact Problem}
	\label{sec:non-unique}

	\begin{figure}
	\resizebox{\linewidth}{!}{
		\includegraphics[width=0.7\textwidth]{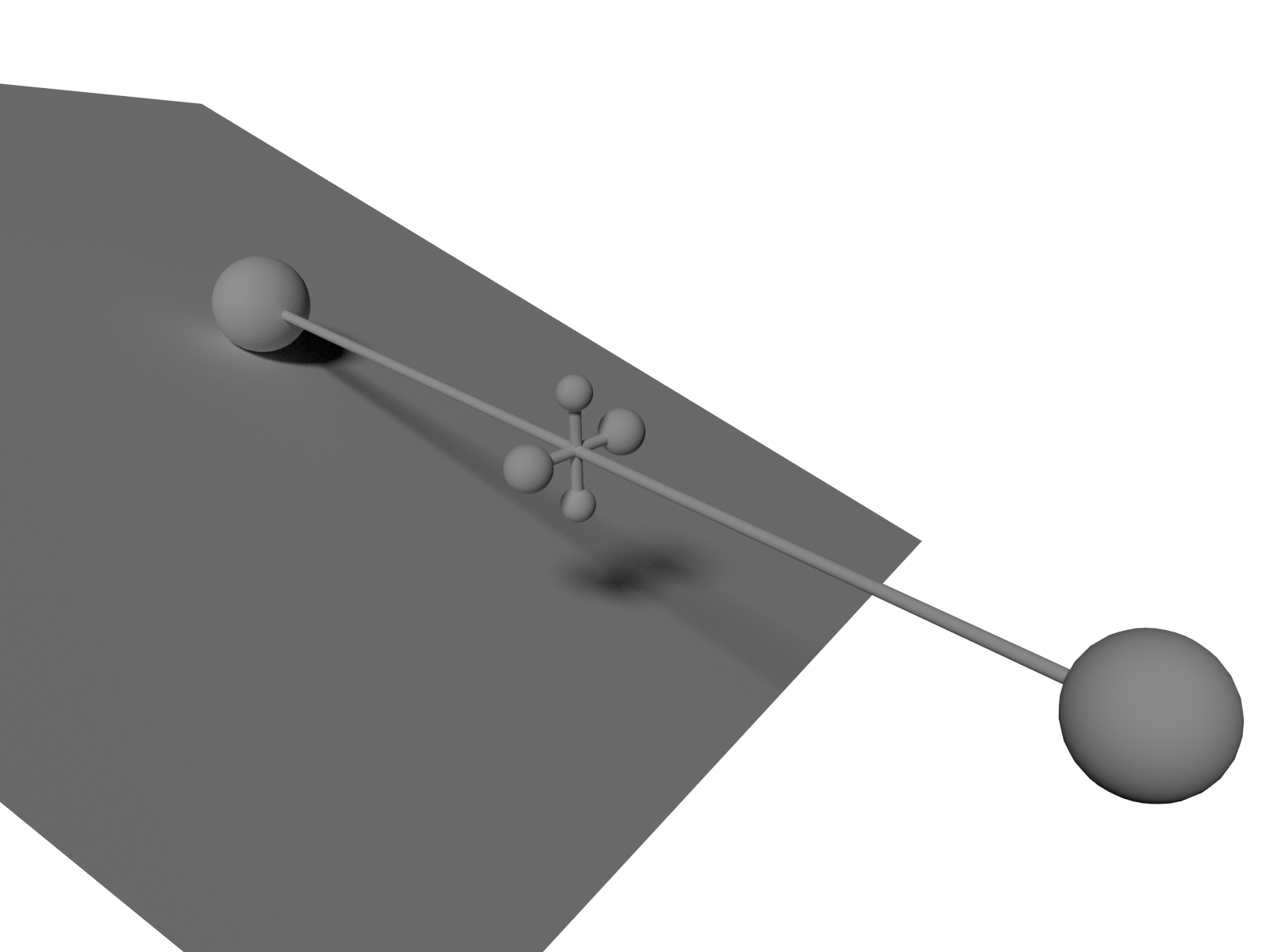}
		\caption{Illustration of the academic single-contact problem. An object constituted out of six point masses symbolized as spheres connected by massless rods collides with an inclined plane. The rods are aligned with the coordinate axis and the point masses are arranged symmetrically on each axis.}
		\label{fig:academic}
	}
	\end{figure}

		As an artificially constructed example consider a rigid body composed of
		six mass points connected by massless rods. The mass points are located
		symmetrically on the axes ($\pm d_x \vec e_x, \pm d_y \vec e_y, \pm d_z \vec e_z$) and both mass points on each axis concentrate
		the same mass ($m_x, m_y, m_z$). The center of mass thus coincides with
		the origin and the inertia tensor is given by
		\begin{equation*}
			\inertia{} = 2\diag(m_y d_y^2 + m_z d_z^2, m_x d_x^2 + m_z d_z^2, m_x d_x^2 + m_y d_y^2).
		\end{equation*}
		Let the mass point in the positive x direction contact a plane with
		normal $\vec n$. Let $\vec t = \vec n \cross \vec e_x$ and
		$\vec o = \vec n \cross \vec t$ span the tangential plane. Let
		$\mat{Q} = \begin{bmatrix}\vec n\,\vec t\,\vec o\end{bmatrix}$ denote the
		contact frame. Then
		\begin{equation*}
			\mat{A} = \mat{Q}^\transp ( (2(m_x + m_y + m_z))^{-1} \identmat{} - d_x^2 \vec e_x^\cross \inertia{}^{-1} \vec e_x^\cross) \mat{Q}.
		\end{equation*}

\begin{figure}
			\includegraphics[width=0.7\textwidth]{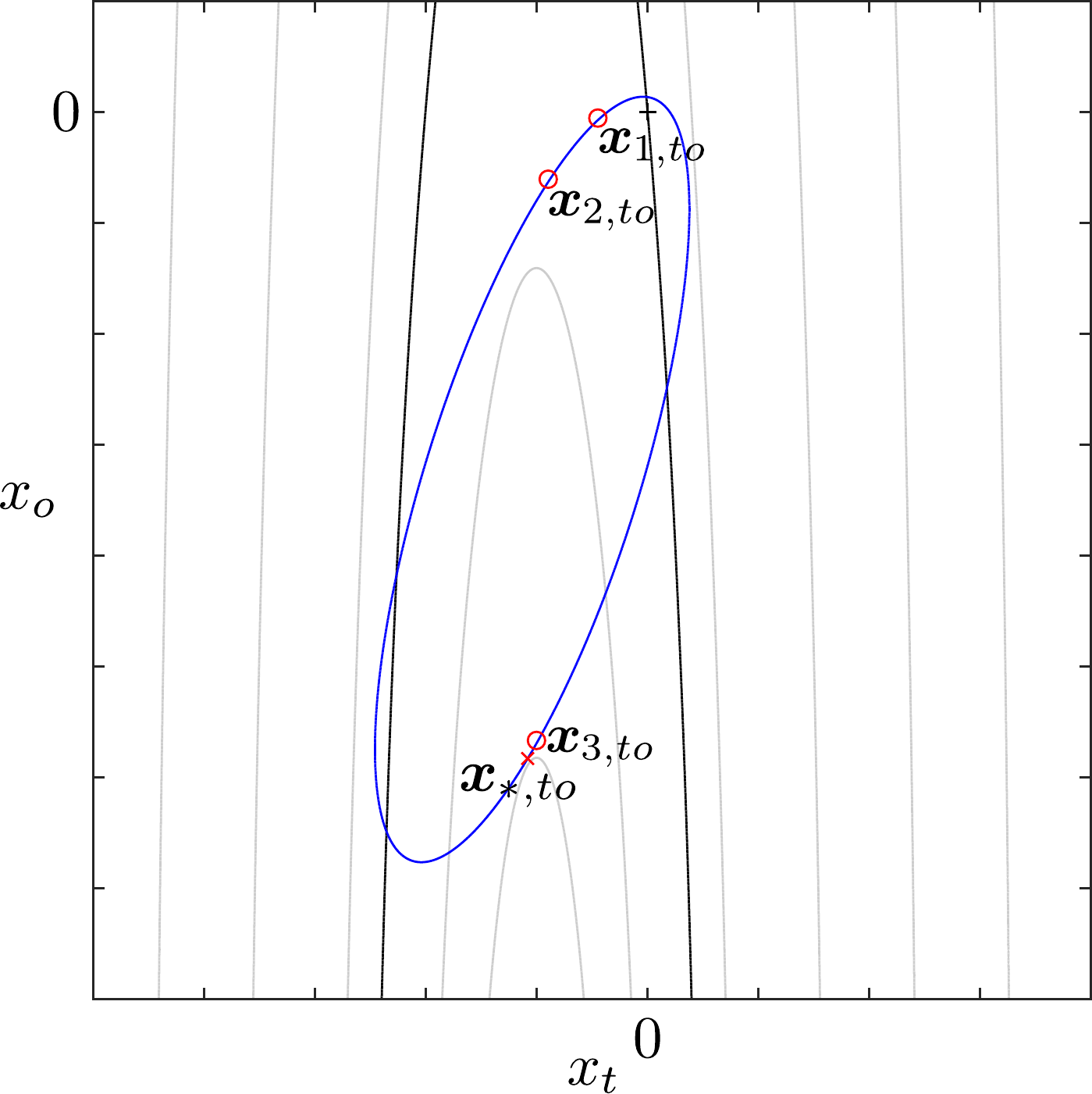}
			\caption{Illustration of the academic contact problem including contour lines of the objective function, constraint set and solutions.}\label{fig:non-unique}
		\end{figure}

		A rendering of this problem using $d_x = 10$, $d_y = d_z = 1$, $m_x = 0.03$, $m_y = 65$,
		$m_z = 50$, $\cof{} = 3.7$,
		$\vec{\hat{n}} = (0.25, 0.36, -0.9)^\transp$, $\vec n = \sfrac{\vec{\hat{n}}}{\norm{\vec{\hat{n}}}_2}$
		and $\vec b = \mat{Q}^\transp (0.06, 0.7, 0.23)^\transp$ is shown in Fig.~\ref{fig:academic}.
		The Coulomb solutions can be calculated for example using the polynomial
		root-finding approach described in \cite{bonnefon11}. The contact
		problem has three dynamic Coulomb solutions $\vec x_i$ ($i \in \{1,2,3\}$) directly opposing
		the post-impulse relative contact velocity in the tangential plane and a unique maximally dissipative solution $\vec x_*$.
		Approximate numerical values are listed in Tab.~\ref{tab:non-unique} for reference.
		\begin{table}
			\resizebox{\linewidth}{!}{
			\begin{tabular}{lll}
				\toprule
				$\approx \vec x_i^\transp$   & $\approx (\mat{A} \vec x_i - \vec b)^\transp$ & $\approx \frac{1}{2}\vec x_i^\transp \mat{A} \vec x_i - \vec x_i^\transp \vec b$ \\
				\midrule
				$\tvec{0.12}{-0.44}{-0.065}$ & $\tvec{0}{0.40}{0.060}  $ & $-0.252$ \\
				$\tvec{0.29}{-0.89}{-0.62} $ & $\tvec{0}{0.082}{0.057} $ & $-0.393$ \\
				$\tvec{1.56}{-0.99}{-5.7}  $ & $\tvec{0}{0.0061}{0.035}$ & $-0.631$ \\
				\bottomrule \toprule
				$\approx \vec x_*^\transp$   & $\approx (\mat{A} \vec x_* - \vec b)^\transp$ & $\approx \frac{1}{2}\vec x_*^\transp \mat{A} \vec x_* - \vec x_*^\transp \vec b$ \\
				\midrule
				$\tvec{1.6}{-1.1}{-5.8}$ & $\tvec{0}{-0.057}{0.034}$ & $-0.634$ \\
				\bottomrule
			\end{tabular}}
			\caption{Approximate Coulomb solutions and maximally dissipative solution for the numerically constructed single-contact problem including relative contact velocities in the contact frame and objective function values.\label{tab:non-unique}}
		\end{table}
		\figref{fig:non-unique} plots the contour lines of the objective function for points in the plane of maximum compression and the
		boundary of the constraint set in solid blue. The red cross marks the maximally dissipative solution along the ellipse. The red circles mark the three Coulomb solutions.
	
	\subsection{Paradox Single-Contact Problem}
	\label{sec:paradox}

	\begin{figure}
	\resizebox{\linewidth}{!}{
		\includegraphics[width=0.7\textwidth]{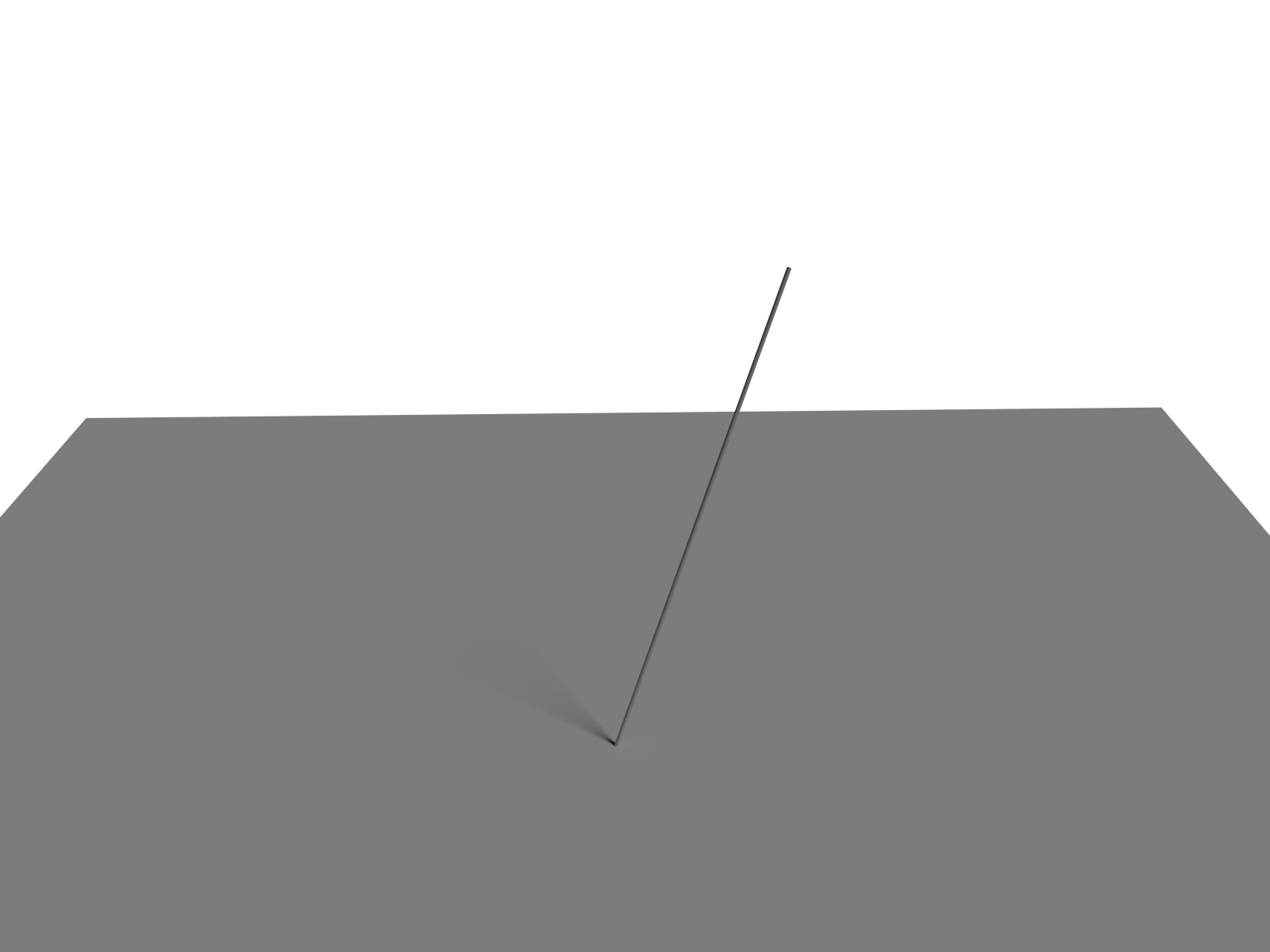}
		\caption{Illustration of the paradox single-contact problem. A slender rod is grazing along a plane.}
		\label{fig:rod}
	}
	\end{figure}

	Let a contact exist at time $t_0$ between a slender rod of length $l \in \R^{>0}$ and a half-space $\setprop{\vec x \in \R^3}{\vec n^\transp \vec x \leq 0}$ with normal $\vec n = \unitvec{y}$. Let $\vec t = \unitvec{x}$ and $\vec o = \unitvec{z}$ complete the contact frame, where $\begin{bmatrix} \unitvec{x} & \unitvec{y} & \unitvec{z} \end{bmatrix} = \identmat{3}$.
	Let $0 < \theta(t_0) = \theta_0 < \frac{\pi}{2}$ denote the initial angle between the plane and the rod. Let the slender rod correspond to rigid body~$a$ and let the half-space correspond to rigid body~$b$.
	The half-space is stationary such that $\linvel{b}(t) = \angvel{b}(t) = \vec 0$, $\mass{b} = \infty$ and $\inertia{b} = \infty \identmat{3}$.
	The rod shall be centered at the origin and aligned along the x axis of the body frame.
	Then, with uniformly distributed mass $\mass{a}$, its body-frame inertia tensor is $\inertia{a,0} = \frac{m l^2}{12} \diag (\infty, 1, 1)^\transp$.
	The rod is located in the x-y plane $\setprop{\vec x \in \R^3}{\unitvec{z}^\transp \vec x = 0}$ of the inertial frame with
	its center of mass at $\pos{a}(t_0) = \frac{l}{2} (\cos \theta_0, \sin \theta_0, 0)^\transp$, such that
	the contact is closed initially~($\xi(t_0) = 0$). Let $\preangvel{a}(t_0) = \vec 0$
	and $\prelinvel{a}(t_0) = -\unitvec{x}v_0$ with $v_0 \in \R^{>0}$, such that initially
	the contact neither separates nor collides ($\dot{\xi}^-(t_0) = \prerelvelCFn{}(t_0) = 0$).
	Let the contact position function $\contactpos{}$ track the lower tip of the slender rod.
	Let gravity $\vec g = -\unitvec{y} g_0$ with $g_0 \in \R^{>0}$ act. A visualization of this setup is displayed in Fig.~\ref{fig:rod}.
	Under these conditions the contact problem is essentially planar and corresponds to the paradox configuration
	published by Painlev\'{e}~\cite{painleve1895,stewart00}.

	If $t_0$ is a non-impulsive point in time ($t_0 \not \in \mathcal{T}_q$), $\dot{\xi}^-(t_0) = \dot{\xi}^+(t_0)$ and
	since $\xi(t_0) = \dot{\xi}^{\sfrac{-}{+}}(t_0) = 0$, the acceleration-level non-penetration constraint
	is enabled from~\figref{fig:contact_constraints}, where
	\begin{equation*}
		\begin{split}
			\ddot{\xi}^{\sfrac{-}{+}}(t_0) & = \prepostrelveldotCFn{}(t_0)\ \text{and}\\
			\prepostrelveldot{}(t_0) & = \prepostlinveldot{a}(t_0) + \prepostangveldot{a}(t_0) \cross (\contactpos{}(t_0) - \pos{a}(t_0)) \\
			                         & + \prepostangvel{a}(t_0) \cross (\prepostangvel{a}(t_0) \cross (\contactpos{}(t_0) - \pos{a}(t_0))) \\
									 & = \mat{W}(t_0)^\transp \dvect{\prepostlinveldot{a}(t_0)}{\prepostangveldot{a}(t_0)} \\
									 & = \mat{W}(t_0)^\transp \mat{M}(\orient{}(t_0))^{-1} \mat{W}(t_0) \contactforce{}(t_0) + \vec g.
		\end{split}
	\end{equation*}
	Since the contact is sliding at time~$t_0$, the frictional contact reaction force is known to be
	\begin{equation*}
		\contactforceCFto{}(t_0) = -\cof{} \contactforceCFn{}(t_0) \frac{\prepostrelvelCFto{}(t_0)}{\norm{\prepostrelvelCFto{}(t_0)}_2} = \cof{} \contactforceCFn{}(t_0) \dvect{1}{0}.
	\end{equation*}
	Consequently,
	\begin{equation*}
		\contactforce{}(t_0) = \begin{bmatrix} \vec n & \vec t & \vec o \end{bmatrix} \tvect{\contactforceCFn{}(t_0)}{\cof{} \contactforceCFn{}(t_0)}{0} = \tvect{\cof{}}{1}{0} \contactforceCFn{}(t_0).
	\end{equation*}
	Since gravity acts, the acceleration-level non-penetration constraint is
	compelled to be active. Thus,
	\begin{equation*}
		\prepostrelveldotCFn{}(t_0) = \vec n^\transp \mat{A}(t_0) \tvect{\cof{}}{1}{0} \contactforceCFn{}(t_0) - g_0 = 0,\ \contactforceCFn{}(t_0) \geq 0,
	\end{equation*}
	where $\mat{A}(t_0)$ is given by Eq.~\eqref{eq:diagblock}:
	\begin{equation*}
		\shrinkeqnnew{1}{
		\mat{A}(t_0) = \mass{a}^{-1} \begin{bmatrix}1 + 3\sin^2 \theta_0 & -3\sin \theta_0 \cos \theta_0 & 0 \\ -3\sin \theta_0 \cos \theta_0 & 1 + 3 \cos^2 \theta_0 & 0 \\ 0 & 0 & 1 + 3(\cos^2 \theta_0 - \sin^2 \theta_0)^2 \end{bmatrix}.
		}
	\end{equation*}
	The coefficient is then
	\begin{equation*}
		\shrinkeqnnew{1}{
		\vec n^\transp \mat{A}(t_0) \tvect{\cof{}}{1}{0} = \mass{a}^{-1} (1 + 3\cos \theta_0 (\cos \theta_0 - \cof{} \sin \theta_0)).
		}
	\end{equation*}
	The equation is not solvable for non-negative $\contactforceCFn{}(t_0)$ if
	the coefficient is negative. The sign of the coefficient depends on $\cof{}$
	and $\theta_0$. Thus, for a given angle $0 < \theta_0 < \frac{\pi}{2}$ the
	contact problem at hand has a non-impulsive solution if
	\begin{equation*}
		\cof{} < \cof{*}(\theta_0) = \frac{\cos \theta_0 + \frac{1}{3\cos \theta_0}}{\sin \theta_0},
	\end{equation*}
	where the lowest bound on $\cof{}$ is $\frac{4}{3}$ at the angle $\frac{1}{2} \cos^{-1} \left(-\frac{3}{5}\right)$.
	If this condition is not met, the assumption that $t_0$ is a non-impulsive point
	in time is wrong. Instead an impact problem has to be solved beforehand.

	The impact model with purely inelastic impacts and Coulomb-like friction as
	presented in \figref{fig:contact_constraints} exhibits multiple solutions.
	The zero solution $\contactimpulse{}(t_0) = \vec 0$ is perfectly valid,
	since no collision is taking place ($\xi(t_0) = \dot{\xi}^-(t_0) = 0$).
	However, it clearly does not lead to a post-impact state with a non-impulsive
	solution.

	Any other solution must be located on the plane of maximum compression
	($\dot{\xi}(t_0) = 0$) and in the friction cone.
	The impulse necessary to obtain a post-impulse sticking contact state is
	determined by the vector equation 
	$\postrelvel{}(t_0) = \mat{A}(t_0) \contactimpulse{0} + \prelinvel{a}(t_0) = \vec 0$:
	\begin{equation*}
		\shrinkeqnnew{1}{
		\begin{split}
			\contactimpulse{0} & = \frac{\mass{a}}{8} \begin{bmatrix} 5 + 3\cos(2 \theta_0) & 3 \sin(2\theta_0) & 0 \\ 3 \sin(2\theta_0) & 5 - 3 \cos(2 \theta_0) & 0 \\ 0 & 0 & \frac{16}{5 + 3 \cos(4 \theta_0)}\end{bmatrix} \begin{pmatrix} v_0 \\ 0 \\ 0 \end{pmatrix} = \\
			                   & = \frac{\mass{a} v_0}{8} (5 + 3\cos(2\theta_0), 3 \sin (2 \theta_0), 0)^\transp.
		\end{split}
		}
	\end{equation*}
	It is easily verified, that $\contactimpulse{0}$ resides within the
	friction cone for any $0 < \theta_0 < \frac{\pi}{2}$ and any
	coefficient of friction requiring an impulsive solution ($\cof{} \geq \cof{*}(\theta_0)$):
	\begin{equation*}
		\begin{split}
			\norm{\contactimpulseCFto{0}}_2 & = \frac{\mass{a} v_0}{8} (5 + 3\cos (2 \theta_0)) \\
			                                & = \cof{*}(\theta_0) \frac{\mass{a} v_0}{8} 3 \sin(2\theta_0) \\
			                                & \leq \cof{} \frac{\mass{a} v_0}{8} 3 \sin(2\theta_0) = \cof{} \contactimpulseCFn{0}.
		\end{split}
	\end{equation*}
	Thus $\contactimpulse{}(t_0) = \contactimpulse{0}$ is a second solution
	of the impact problem effecting a slip-stick transition. In fact,
	there also can be an infinite number of sliding solutions. For example in
	the edge case where $\cof{} = \cof{*}(\theta_0)$, all convex combinations
	of the $\vec 0$ solution and the $\contactimpulse{0}$ solution are
	also solutions, where the slip directly opposes the frictional impulse.

	When replacing the impact model with the Coulomb-like friction by
	the impact model complying with the maximum dissipation principle,
	the structogram in \figref{fig:structogram} easily identifies the
	unique impulsive solution: If $\cof{} \geq \cof{*}(\theta_0)$, then
	$\contactimpulse{}(t_0) = \contactimpulse{0}$.
	It is the same impulsive solution as the sticking solution in
	the impact model with the Coulomb-like friction. Conversely, if
	a non-impulsive solution exists ($\cof{} < \cof{*}(\theta_0)$), then
	$\contactimpulse{0}$ is not contained in the friction cone and
	predicate $P_2$ in the structogram then always selects $\contactimpulse{}(t_0) = \vec 0$:
	\begin{equation*}
		\begin{split}
			A_{nn} & = 1 + 3 \cos^2 \theta_0 = \cof{*}(\theta_0) 3 \sin \theta_0 \cos \theta_0 \\
			       & > \cof{} 3 \sin \theta_0 \cos \theta_0 = \cof{} \sqrt{A_{nt}^2 + A_{no}^2}.
		\end{split}
	\end{equation*}
	Thus the impact problem results in the zero solution if a non-impulsive
	solution exists as expected. The impact model with Coulomb-like friction
	behaves the same way in this respect: If $\cof{} < \cof{*}(\theta_0)$ then the plane of
	maximum compression intersects the friction cone only at $\vec 0$.

	The impulsive contact reaction $\contactimpulse{}(t_0) = \contactimpulse{0}$
	leads in both impact models to a sticking post-impulse contact state,
	where non-impulsive contact reactions exist for the subsequent contact problem.
	However, alternate solutions of the impact model with Coulomb-like friction
	do not necessarily lead to subsequent contact problems with non-impulsive
	solutions.

	\subsection{Macro-Scale Behaviour}

	The macro-scale behaviour of the proposed maximum dissipation friction model is compared to the Coulomb friction model by numerical simulations of fast granular channel flows. All simulations are carried out with the pe module of the waLBerla multi physics framework freely available at \emph{walberla.net}. The algorithms are based on time stepping methods presented in Sec.~\ref{sec:numerics}. The parallel implementation is described in \cite{preclik14,preclik15}. For the computations the compute resources of the Regionales Rechenzentrum Erlangen (RRZE) are used. The Emmy cluster comprises 560 compute nodes, each equipped with two Xeon 2660v2 proccessors (10 cores, 2-way SMT) clocked at 2.2\,GHz and 64\,GiB of RAM.

	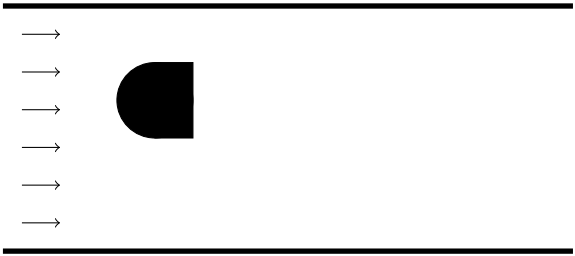
\begin{figure}
		\begin{tikzpicture}[scale=0.05]
			\draw [-, line width=2pt] (0,0) -- (150,0);
			\draw [-, line width=2pt] (0,65) -- (150,65);

			\draw [fill=black] (40,30) rectangle (50,50);
			\draw [fill=black] (40,40) circle [radius=10];

			\draw [->] (5,10 - 2.5) -- (15, 10 - 2.5); 
			\draw [->] (5,20 - 2.5) -- (15, 20 - 2.5); 
			\draw [->] (5,30 - 2.5) -- (15, 30 - 2.5); 
			\draw [->] (5,40 - 2.5) -- (15, 40 - 2.5); 
			\draw [->] (5,50 - 2.5) -- (15, 50 - 2.5); 
			\draw [->] (5,60 - 2.5) -- (15, 60 - 2.5); 
		\end{tikzpicture}
		\caption{Top-down view of the simulation area. The particle channel is confined by solid walls in y and z direction. A constant particle inflow at the left side is artificially created. Within the channel an obstacle is present to disturb the particle flow.}
		\label{fig:SimulationArea}
	\end{figure}

	The simulation domain is a rectangular channel confined by solid walls in the y and z direction. Its dimensions are $\SI{15}{cm} \times \SI{6.5}{cm} \times \SI{2.0}{cm}$. The channel is filled with monodisperse spherical particles with a diameter of \SI{0.47}{mm} and a density of \SI{2.65}{g/cm^3}. These are arranged on a regular rectangular grid with a spacing of \SI{1}{mm}. All particles are given an initial velocity of \SI{1}{m/s} in positive x direction. A random perturbation velocity in y and z direction is applied with each component varying between \SI{-0.5}{m/s} and \SI{0.5}{m/s}. To produce a steady inflow a moving plane with infinite mass is added at the left end of the channel ($x=\SI{0}{cm}$). This plane moves with \SI{1}{m/s} pushing the particles into the channel. After the plane has moved a distance equal to two times the radius of the particles the position of the plane is reset and a new layer of particles is generated. This process is repeated throughout the whole simulation generating an inflow rate of roughly \num{1.4e6} particles per second. The other end of the channel remains open and the particles leaving the channel are deleted. All the particles are also influenced by a gravitational acceleration of \SI{9.81}{m/s^2} in negative z direction. In the region $[\SI{3}{cm}, \SI{5}{cm}] \times [\SI{3}{cm}, \SI{5}{cm}] \times [\SI{0}{cm}, \SI{2}{cm}]$ a stationary object obstructing the channel is introduced. The obstacle is composed of a cylinder with radius \SI{1}{cm} located at (\SI{4}{cm}, \SI{4}{cm}, \SI{1}{cm}) with its axis aligned along the z axis. The right half of the cylinder is replaced by a box. A cross-section of this setup is sketched in Fig.~\ref{fig:SimulationArea}.

	For the computation the domain is decomposed into $16 \times 10 \times 4$ evenly sized rectangular subdomains. Each subdomain is handled by one process totaling to 640 processes which are distributed onto 32 nodes of the Emmy cluster. The time-step length is chosen to be \SI{10}{\micro s}. This guarantees a stable simulation. The total simulation time is \SI{10}{s} resulting in \num[retain-unity-mantissa = false]{1e6} time steps. After these time steps all initial perturbations are resolved. The implementation of the collision resolution solver described in \cite{preclik14} is used. Both friction models use a coefficient of friction equal to~\num{0.4}.

	\begin{table}
	\begin{minipage}[t][2.5cm][t]{\linewidth}
	\vfill
	\resizebox{\linewidth}{!}{
	\begin{tabular}{lrr}
		\toprule
		                           & Coulomb     & Max. Dissipation \\
		\midrule
		avg. number of particles                & \num[scientific-notation=true, round-mode = figures, round-precision=3]{246992.2917082917} & \num[scientific-notation=true, round-mode = figures, round-precision=3]{247005.05094905096} \\
		solid volume fraction      & \SI{57.2}{\percent} & \SI{57.2}{\percent} \\ 
		avg. number of contacts                 & \num[scientific-notation=true, round-mode = figures, round-precision=3]{320276.5074925075} & \num[scientific-notation=true, round-mode = figures, round-precision=3]{320196.8191808192} \\
		avg. number of contacts per particle    & \num[round-mode = figures, round-precision=3]{2.593412978820955} & \num[round-mode = figures, round-precision=3]{2.592633777734896} \\
		max. penetration (\si{\micro m})  & \num[round-mode = figures, round-precision=3]{39.0600228} & \num[round-mode = figures, round-precision=3]{46.49761679} \\
		\bottomrule
	\end{tabular}}
	\vfill
	\end{minipage}
	\caption{Key figures collected for both simulations. Samples were collected every \SI{10}{ms} during the \SI{10}{s} total simulation time. }
	\label{tab:statistics}
	\end{table}

	Some general information about the simulation is collected every 1000 time steps and averaged over the complete simulation. Within the range of accuracy both friction models result in almost identical values which are summarized in Tab. \ref{tab:statistics}. The average number of particles in the simulation domain is \num{2.47e5} which leads to a solid volume fraction of \SI{57.2}{\percent}. During every time step an average of \num{3.20e5} contacts which equals \num{2.59} contacts per particle are resolved. The maximum penetration depth between two particles was \SI[round-mode = figures, round-precision=3]{39.0600228}{\micro m} (\SI{8.31}{\percent} of the particle radius) in the Coulomb friction case and \SI[round-mode = figures, round-precision=3]{46.49761679}{\micro m} (\SI{9.89}{\percent} of the particle radius) in the maximum dissipation case.

	\begin{figure}
	\resizebox{\linewidth}{!}{
		\includegraphics[width=0.7\textwidth]{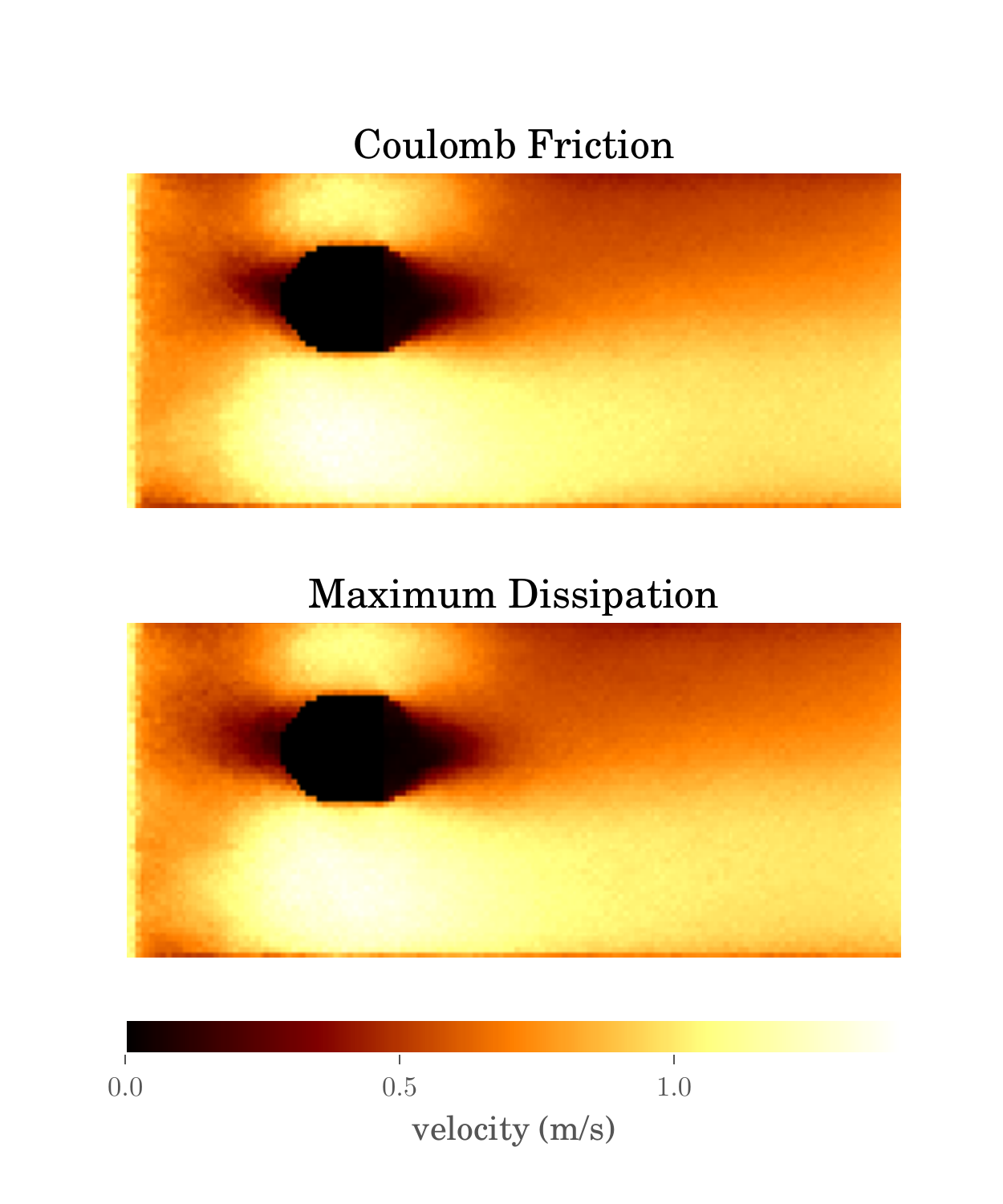}
		\caption{Velocity profile of the particles throughout the channel. Around the obstacle the velocity of the particles increases due to the narrowing of the channel. After the obstacle collisions as well as friction slow down the fast particles again.}
		\label{fig:VelComparison}
	}
	\end{figure}

	\begin{figure}
	\resizebox{\linewidth}{!}{
		\includegraphics[width=0.7\textwidth]{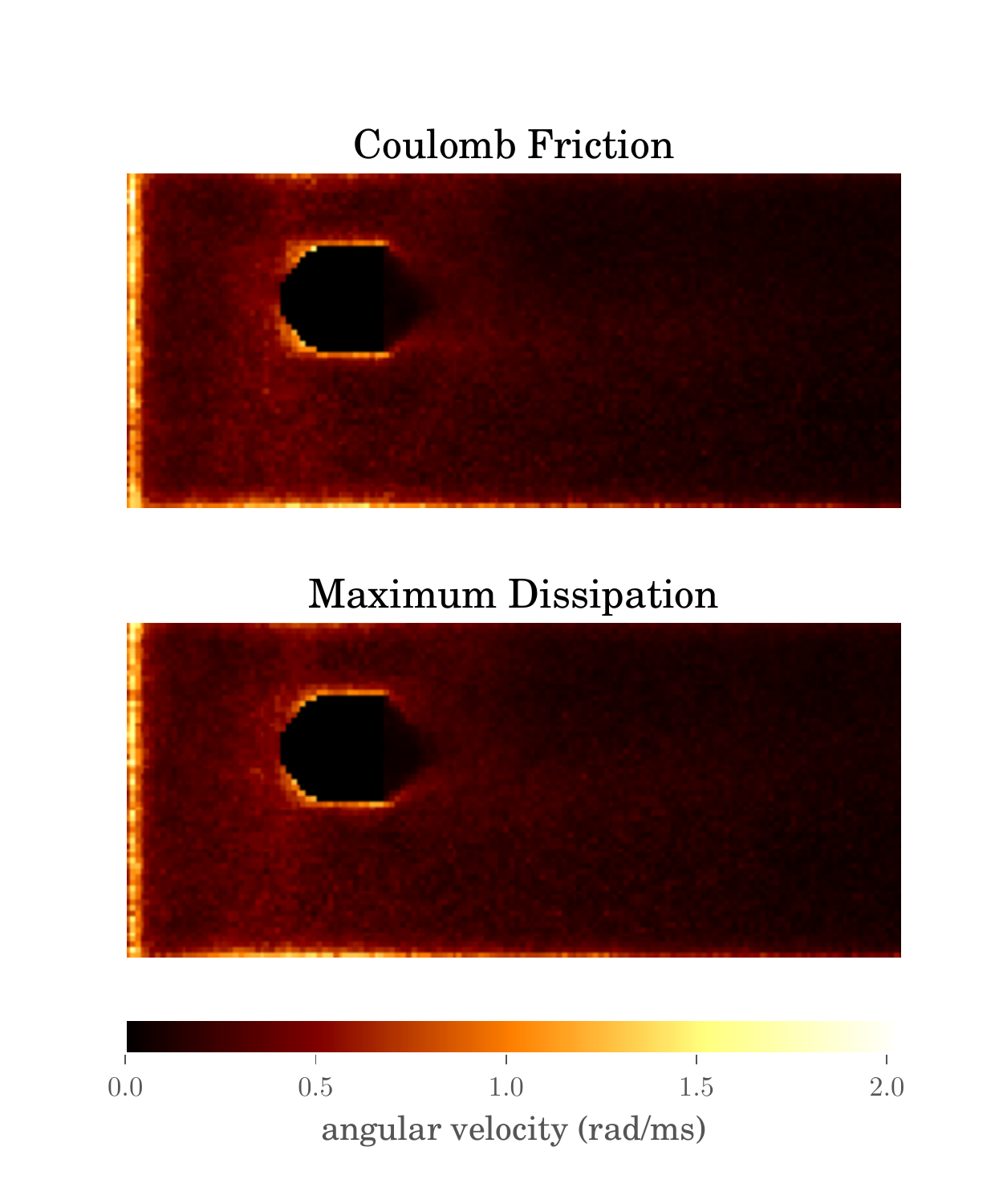}
		\caption{Angular velocity profile of the particles throughout the channel. One can see that rotations do not play a major role in this setup as friction damps any angular velocity rapidly. A certain amount of rotating particles can be found at the inflow and near the stationary object in the middle of the channel.}
		\label{fig:AngVelComparison}
	}
	\end{figure}

	To validate the accordance of both friction models the last frame of each simulation is used and the particle configuration is analyzed. The particles are sorted into $\num{150} \times \num{65}$ equally sized cells in the x-y plane ignoring the z axis and the velocity (see Fig. \ref{fig:VelComparison}) as well as the angular velocity (see Fig. \ref{fig:AngVelComparison}) is averaged over all particles within one cell. Both simulations show a very similar velocity profile throughout the whole channel. The particles right after the inflow have very high velocities but are damped rapidly. After a short almost homogeneous region the particles gain speed again as the channel narrows around the obstacle. The particle velocities peak at \SI{1.42}{m/s} (Coulomb Friction) and \SI{1.39}{m/s} (Maximum Dissipation) respectively. The angular velocity comparison shows a similar picture. After a short initial region the angular velocity gets damped heavily. At the boundary of the stationary obstacle and the walls, the angular velocity stays high due to friction. Both friction models result visually in a similar behaviour.

	\begin{figure}
		\includegraphics[width=1.0\textwidth]{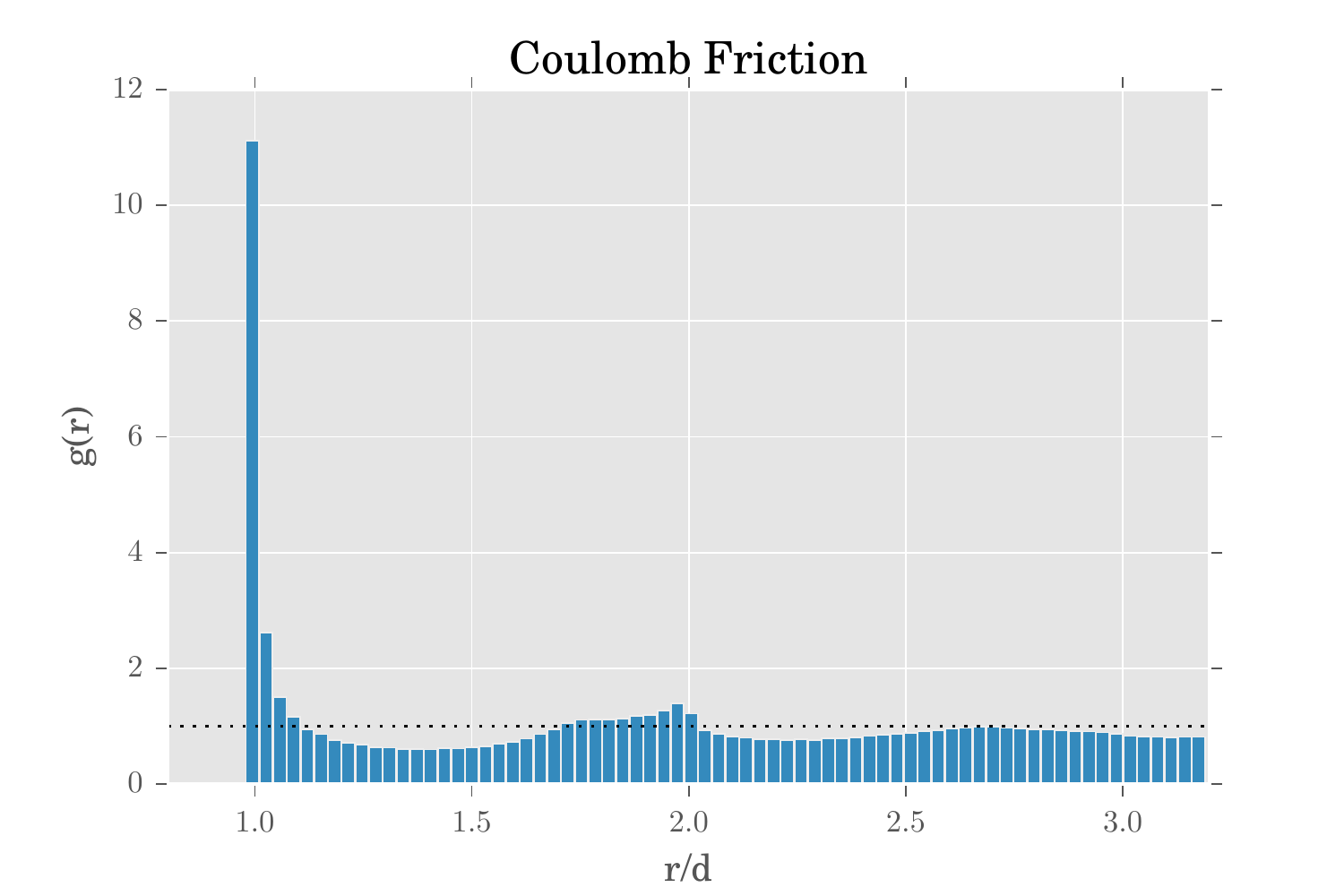}
		\includegraphics[width=1.0\textwidth]{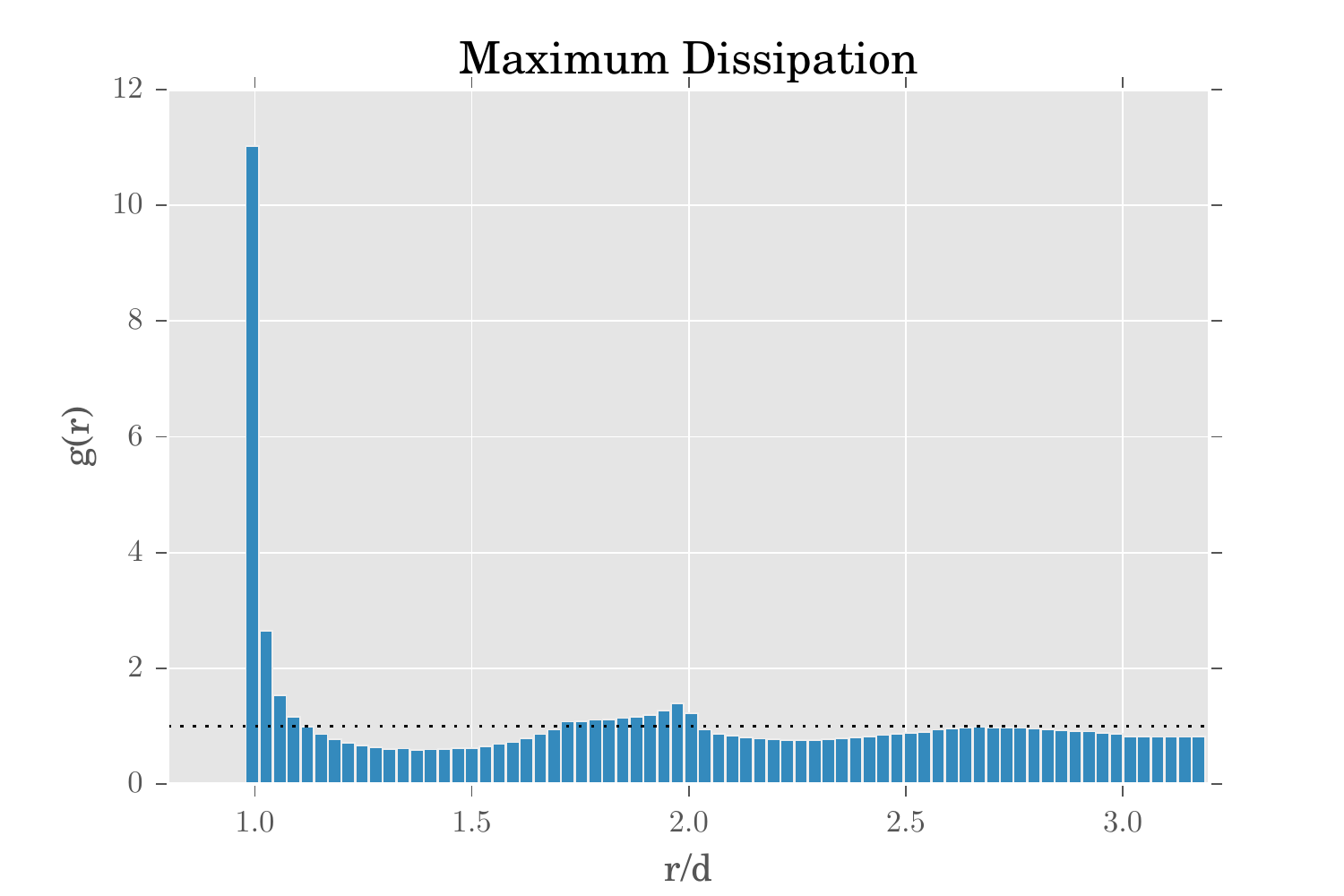}
		\caption{Radial distribution function of the particles near the outflow. The radial distance $r$ is given in units of the particle diameter $d=\SI{0.47}{mm}$. The dotted line $g(r) = 1$ represents the value for a completely amorphous material.}
		\label{fig:RDF}
	\end{figure}
	
	As a last check the radial distribution function of the particles near the exit is analyzed to obtain insight into the spatial arrangement. The region $[\SI{12}{cm}, \SI{14}{cm}] \times [\SI{0}{cm}, \SI{6.5}{cm}] \times [\SI{0}{cm}, \SI{2}{cm}]$ is chosen. All the inter-particle distances are calculated and summed up in a histogram. The histogram is then normalized by the total number of particles and the expected amount of particles within each individual bin - $\int_{x_0}^{x_1}4\pi\rho r^2 \text{d}r$, with $\rho$ being the overall particle density and $x_0$, $x_1$ being the boundaries of the corresponding histogram bin. For both friction models the histogram can be found in Fig. \ref{fig:RDF}. They match almost exactly. One can see a strong peak at a distance equal to the particle diameter which indicates a dense packing. The rest of the histogram does not reveal additional favored distances suggesting an amorphous packing.

	The numerical experiment described above shows no significant difference between the two friction models. The newly developed maximum dissipation friction model thus reproduces the expected macro scale behaviour.

	\section{Summary}
	\label{sec:summary}

		In this paper, an alternative purely inelastic frictional impact model is
		presented. In that impact model the contact reaction impulses consistently
		maximize dissipation leading to unique contact reactions for single-contact
		impact problems. An academic single-contact impact problem is analyzed, demonstrating
		the potential non-uniqueness of the impulsive contact reactions if
		Coulomb-like friction constraints act instead. Furthermore, a paradox
		single-contact impact problem was analyzed. It was found that the impact
		model maximizing dissipation results in the solution producing a slip-stick
		transition, whereas the impact model with Coulomb-like friction in addition allows
		the zero solution and possibly even an infinite number of dynamic solutions,
		which not necessarily result in configurations with subsequent
		non-impulsive solutions. The paper also shows how the
		impact model based on the maximum dissipation principle can be embedded
		in an impulse-velocity
		time-stepping scheme. A numerical experiment is conducted to demonstrate
		that changing from an impact model with Coulomb-like friction to the maximally
		dissipative impact model in the time-stepping scheme has a negligible
		influence on the macroscopic behaviour of a granular channel flow past an obstacle.
		The multi-contact problem is solved
		using a blend between a non-linear block Gauss-Seidel and a weighted
		non-linear block Jacobi, where each block corresponds to a single-contact
		problem. The paper presents an analytic solution of the single-contact
		problem that can act as a subsystem solver in the non-linear block
		relaxation method. The analytic solution involves transforming the
		single-contact problem into a quartic equation with complex coefficients.
		Back transformation and filtering of the solutions of the quartic equation
		leads to the unique contact reaction maximizing dissipation. Additionally, the
		analytic solution can be used to resolve two-particle collisions in event-driven
		integrations of rigid-body dynamics.

		The presented impact model is missing a restitution hypothesis for
		partly elastic impacts. Even though applying Poisson's hypothesis is
		straightforward, the extension of the impact model by an energetically consistent restitution hypothesis
		is not. Also proving or falsifying that the constructed time-stepping scheme
		converges to a solution of the corresponding integral equations remains an open
		problem. Besides addressing the non-uniqueness in the impact model, the
		approach might also be transferable to remove non-uniqueness in the
		non-compliant contact model when non-penetration and friction constraints
		on the acceleration-level are enabled.

	\section*{Acknowledgment}
		The authors would like to acknowledge the support through the Cluster of Excellence Engineering of Advanced Materials (EAM).

	\appendix

	\section{Appendix}
	\label{sec:appendix}

		\begin{description}
			\item[$c_0 =$]       $\frac{3}{2} \cof{} A_{nn} (A_{no} b_t - A_{nt} b_o)$
			\item[$c_1 =$]       $(\cof{}^4 ((A_{no}^2 (A_{tt}^2 + A_{to}^2) - 2 A_{nt} A_{no} A_{to}\allowbreak (A_{tt} + A_{oo}) + A_{nt}^2 (A_{to}^2 + A_{oo}^2)) b_n^2 + 2 (A_{nt} b_o - A_{no} b_t) (A_{no}^2 A_{to} + A_{nt}\allowbreak A_{no}\allowbreak (A_{tt} - A_{oo}) - A_{nt}^2 A_{to}) b_n + (A_{nt}^2 + A_{no}^2) (A_{nt} b_o - A_{no} b_t)^2)\allowbreak + 2 \cof{}^2 A_{nn}\allowbreak ((2 A_{nt} A_{no} A_{to} - A_{no}^2 A_{tt} - A_{nt}^2 A_{oo})\allowbreak b_n^2 + A_{nn}\allowbreak (A_{no}\allowbreak (A_{tt} b_o - A_{to} b_t) - A_{nt}\allowbreak (A_{to} b_o - A_{oo} b_t)) b_n + A_{nn}\allowbreak (A_{nt} b_o - A_{no} b_t)^2) + A_{nn}^2 ((A_{nt}^2 + A_{no}^2) b_n^2 - 2 A_{nn}\allowbreak (A_{nt} b_t + A_{no} b_o) b_n + A_{nn}^2 (b_t^2 + b_o^2)))^{\sfrac{1}{2}}$
			\item[$c_2 =$]       $\cof{} A_{nn}\allowbreak ((\frac{1}{4}\allowbreak (A_{tt}^2 + A_{oo}^2) + A_{to}^2 - \frac{1}{2} A_{tt} A_{oo}) b_n^2 + (A_{no}\allowbreak (\frac{1}{2}\allowbreak (A_{tt} - A_{oo}) b_o - A_{to} b_t) - A_{nt}\allowbreak (A_{to} b_o + \frac{1}{2}\allowbreak (A_{tt} - A_{oo}) b_t))\allowbreak b_n + \frac{1}{4}\allowbreak ((A_{nt}^2 + A_{no}^2) b_o^2 + (A_{nt}^2 + A_{no}^2) b_t^2))^{\sfrac{1}{2}}$
			\item[$\varphi_1 =$] $\atan2(\cof{}^2 ((A_{no} A_{to} - A_{nt} A_{oo}) b_n - A_{no}^2 b_t + A_{nt} A_{no}\allowbreak b_o) - A_{nn}^2 b_t + A_{nn} A_{nt} b_n,\allowbreak \cof{}^2 ((A_{nt} A_{to} - A_{no} A_{tt}) b_n + A_{nt} A_{no} b_t - A_{nt}^2 b_o) - A_{nn}^2 b_o + A_{nn} A_{no} b_n)$
			\item[$\varphi_2 =$] $\atan2((A_{tt} - A_{oo}) b_n - A_{nt} b_t + A_{no} b_o,\allowbreak 2 A_{to} b_n - A_{no}\allowbreak b_t - A_{nt} b_o)$
		\end{description}
	
	\printbibliography

\end{document}